\documentclass{ar-1col}

\usepackage[comma]{natbib}
\usepackage{url}
\usepackage{amssymb}
\usepackage{longtable}
\usepackage{lscape}
\usepackage{amsmath}

\jname{Xxxx. Xxx. Xxx. Xxx.}
\jvol{AA}
\jyear{YYYY}
\doi{10.1146/((please add article doi))}

\newcommand{\nion}{\dot{n}_{\mathrm{ion}}}
\newcommand{\fesc}{f_{\mathrm{esc}}}
\newcommand{\Lion}{L_{\mathrm{ion}}}
\newcommand{\Lstar}{L_{\star}}
\newcommand{\Msun}{M_{\odot}}
\newcommand{\Q}{Q_{\mathrm{HII}}}
\newcommand{\rhoUV}{\rho_{UV}}
\newcommand{\rhoSFR}{\dot{\rho}_{\star}}
\newcommand{\um}{\mu\mathrm{m}}
\newcommand{\xiion}{\xi_{\mathrm{ion}}}
\newcommand{\Zsun}{Z_{\odot}}
\newcommand{\MUV}{M_{\mathrm{UV}}}
\newcommand{\LUV}{L_{\mathrm{UV}}}
\newcommand{\phistar}{\phi^\star}
\newcommand{\Mstar}{M^\star}
\newcommand{\mAA}{\text{\normalfont\AA}}
\newcommand{\xHI}{x_{\mathrm{HI}}}
\newcommand{\tage}{t_{\mathrm{age}}}
\newcommand{\mmax}{m_{\mathrm{max}}}
\newcommand{\trec}{\bar{t}_{\mathrm{rec}}}
\newcommand{\Lya}{Ly$\alpha$}
\newcommand{\Lyb}{Ly$\beta$}
\newcommand{\Lyc}{LyC}
\newcommand{\jf}[1]{{\it #1}}
\newcommand{\jp}[1]{{#1}}
\newcommand{\mab}{m_{\mathrm{AB}}}
\newcommand{\texp}{t_{\mathrm{exp}}}
\newcommand{\sqarcmin}{arcmin$^{2}$}

\newcommand{\JWST}{\emph{JWST}}
\newcommand{\HST}{{\it HST}}
\newcommand{\Spitzer}{{\it Spitzer}}
\newcommand{\ALMA}{{\it ALMA}}
\newcommand{\Keck}{{\it Keck}}
\newcommand{\VLT}{{\it VLT}}
\newcommand{\WFC}{{\it WFC3}}

\newcommand{\reion}{Cosmic Reionization}

\newcommand{\nat}{Nature}
\newcommand{\mnras}{Mon. Not. Royal Ast. Soc.}
\newcommand{\aj}{Astron. J.}
\newcommand{\apj}{Astrophys. J.}
\newcommand{\apjl}{Astrophys. J. Lett.}
\newcommand{\apjs}{Astrophys. J. Supp. Series}
\newcommand{\aap}{Astron. \& Astrophys.}
\newcommand{\pasj}{Proc. Astron. Soc. Japan}
\newcommand{\pasa}{Proc. Astron. Soc. Aust.}
\newcommand{\pasp}{Proc. Astron. Soc. Pacific}
\newcommand{\araa}{Ann. Rev. Astron. \& Astrophys.}

\newcommand{\physrep}{Phys. Rep.}
\newcommand{\baas}{Bul. Amer. Astron. Soc.}

\usepackage{hyperref}
\hypersetup{colorlinks=false,linkcolor=blue,filecolor=magenta,urlcolor=cyan}

\begin{document}

\markboth{Robertson}{Galaxies and Reionization before {\it JWST}}


\title{Galaxy Formation and Reionization: Key Unknowns and Expected
Breakthroughs by the {\it James Webb Space Telescope}}

\author{Brant E. Robertson$^1$
\affil{$^1$Department of Astronomy and Astrophysics, University of California, Santa Cruz,
Santa Cruz, CA, USA, 95064; email: brant@ucsc.edu}}

\begin{abstract}
The scheduled launch of {\it James Webb Space Telescope} (\JWST{}) in late 2021 marks a new start 
for studies of galaxy formation at high redshift ($z\gtrsim6$) during the era of Cosmic Reionization.
\JWST{} can capture sensitive, high-resolution images and multi-object 
spectroscopy in the infrared that will transform our view of galaxy formation during the 
first billion years of cosmic history. This review summarizes our current knowledge 
of the role of galaxies in reionizing intergalactic hydrogen ahead of \JWST{}, achieved through 
observations with {\it Hubble Space Telescope} and ground-based facilities including {\it Keck}, 
the {\it Very Large Telescope}, {\it Subaru}, and the
{\it Atacama Large Millimeter/Submillimeter Array}. We identify outstanding questions in
the field that \JWST{} can address during its mission lifetime,
including with the planned \JWST{}
Cycle 1 programs. These findings include:

\vspace{.3cm}
\hangindent=.3cm$\bullet$ Surveys with \JWST{} have sufficient sensitivity and area to complete the census of galaxy formation at the current redshift frontier ($z\sim8$-$10$).

\hangindent=.3cm$\bullet$ Rest-frame optical spectroscopy with \JWST{} of galaxies will newly enable measures of star formation rate, metallicity, and ionization at $z\sim8-9$, allowing for the astrophysics of early galaxies to be constrained.

\hangindent=.3cm$\bullet$ The presence of evolved stellar populations at $z\sim8$-$10$ can be definitively tested by \JWST{}, which would provide evidence of star\\formation out to $z\sim15$.

\end{abstract}

\begin{keywords}
galaxy formation, reionization, structure formation
\end{keywords}
\maketitle

\tableofcontents

\section{INTRODUCTION}
\label{sec:intro}

Ever deeper observations continue to reveal the earliest epochs of galaxy formation.
Searches for distant galaxies now probe 
the era of ``\reion{}'' when hydrogen in the intergalactic medium transitioned from neutral to ionized,
and reach well past its conclusion at redshift $z\sim6$ about 1 billion years after the Big Bang.
By discovering galaxies in the first few hundred million years of cosmic time and measuring their
properties, we can study the connection between galaxy formation and \reion{}. With the 
anticipated launch of {\it James Webb Space Telescope} (\JWST{}) in late 2021, observations of
galaxies during \reion{} will advance dramatically. This review surveys present progress in relating
galaxies with \reion{}, and previews discoveries with \JWST{} that will transform our understanding 
of how early galaxies form and affect intergalactic hydrogen.

To date,
sensitive observations with {\it Hubble Space Telescope} (\HST{}) have detected candidate galaxies
photometrically with plausible redshifts as high as $z\sim11$. The
most distant galaxies with spectroscopically confirmed redshifts lie at redshifts $z\sim9-10$, with
spectra taken by \Keck{}, the {\it Very Large Telescope} (\VLT{}), and the {\it Atacama Large
Millimeter/submillimeter Array} (\ALMA{}). At slightly lower redshifts $z\sim7-9$, photometric
surveys provide $\sim1000$ galaxy candidates to date. Of these, direct spectroscopic confirmation
exists for $\sim100$ at $z\sim7$ and a handful at $z>8$. The majority of these photometric and
spectroscopic discoveries occurred over the last decade since the installation of the \HST{} {\it Wide Field Camera 3} (\WFC{}) in May 2009, as the new sensitivity, field of view, filter array,
and resolution
of \WFC{} relative to its {\it Near Infrared Camera and Multi-Object Spectrometer} predecessor
improved the capabilities of \HST{} to detect and select high-redshift galaxies.
The combination of \HST{} and {\it Spitzer Space Telescope} enabled the first simultaneous
measurements of the rest-frame ultraviolet (UV) and optical spectral energy distributions (SEDs)
of distant galaxies, allowing for joint inferences of the star formation rates (SFRs) and stellar 
masses of early-forming systems. Using the arsenal of observational facilities available to date,
we have inferred the SFR, stellar mass, gas content, metallicity, and dust content in collections
of galaxies at redshifts $z>6$.
With the new capabilities afforded by \JWST{}, the search for the most distant galaxies
will again be reinvigorated. Relative to \HST{} or \Spitzer{}, \JWST{} features a much larger 
collecting area, instruments sensitive to wavelengths $\lambda\sim1-28\um$, and multiplexed
spectroscopic capabilities. \JWST{} will provide a completely new view on \reion{}.

\begin{figure}[h]
\includegraphics[width=\linewidth]{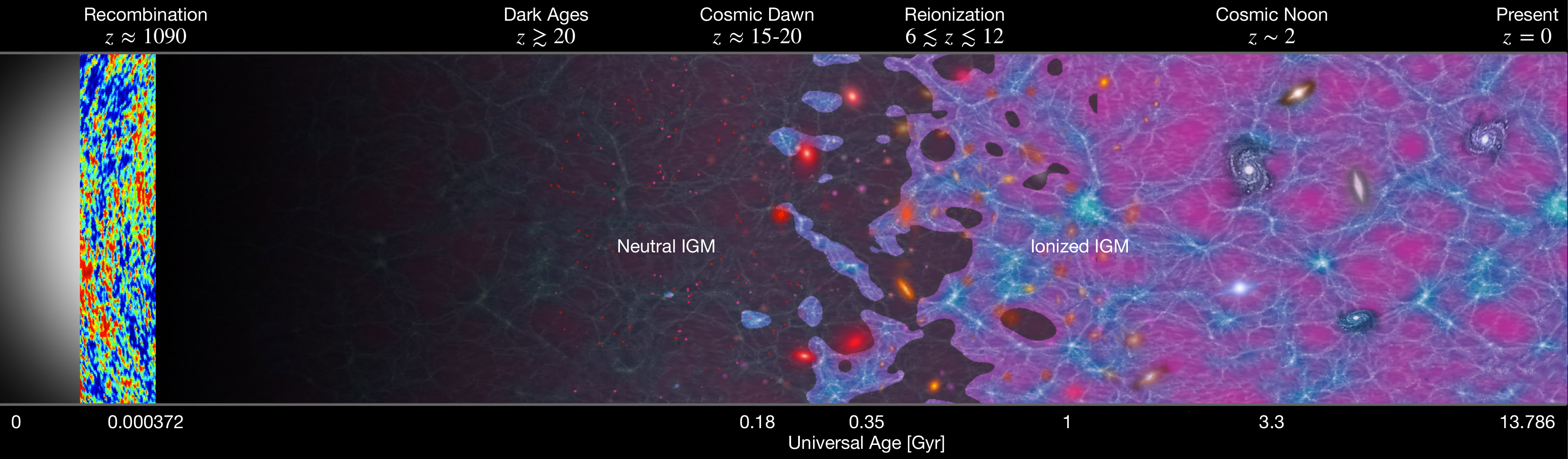}
\caption{Overview of universal history highlighting the epoch of reionization. After the Big Bang ($t=0$), the universe expanded and cooled until electrons could recombine with protons to form neutral hydrogen ($z\approx1090$ at $t\approx372,000$ yrs). During the subsequent Dark Ages, the universe remained dim and neutral until the first stars formed at Cosmic Dawn, with substantial galaxies
forming likely after $z\sim20$ ($t\approx180$ Myr) and perhaps as late as $z\sim15$. 
Once galaxies grew in abundance and luminosity, such that their escaping
Lyman continuum radiation could begin to ionize the surrounding intergalactic
medium (IGM), the process of \reion{} commenced.
Over the period of redshifts $12\lesssim z \lesssim6$, corresponding to times $0.35\lesssim t \lesssim 1$ Gyr, hydrogen in the bulk IGM transitioned from a nearly completely
neutral to a nearly completely ionized state. 
The figure illustrates the changing ionization state of the IGM
as a patchy transition between shaded and unshaded regions, with the 
corresponding evolution
of the galaxy populations represented with simple model images.
Afterward \reion{}, star light and radiation
from active galactic nuclei maintained the ionization of the IGM through the peak
of cosmic star formation at Cosmic Noon ($z\sim2$ and $t\approx3.3$ Gyr) to
the Present ($z=0$ at $t=13.786$ Gyr). This correspondence between redshift and time assumes the \citet{planck2020a} best-fit cosmology. Figure adapted from \citet{robertson2010a}.}
\label{fig:reionization}
\end{figure}

The study of high-redshift galaxies often connects with the novelty of 
finding the most distant or earliest-forming systems. Such discoveries
provide mileposts in our exploration of the nascent universe, where
observationally we have established the presence of virialized structures that
managed to host star formation. Given the technical challenge of detecting faint
distant objects, measuring their fluxes photometrically, and using spectroscopy to
constrain the finer details of their emission, the process of discovering the
earliest galaxies requires a substantial scientific endeavor with its own value. Beyond their 
discovery, the characteristics of individual objects provide a wealth of information about the 
physics of galaxy evolution in the context of cosmological structure formation can be inferred.
Figure \ref{fig:reionization} provides an illustrative diagram of the history
of galaxy evolution, highlighting the epoch of \reion{} at redshifts $z\sim6-12$. In addition to the
many discoveries \JWST{} will provide for galaxies during the last 13 billion years of
cosmic time, \JWST{} can push our knowledge back beyond the height of \reion{} to the
eras when the very earliest galaxy forms.
Given this context for discoveries, the forthcoming launch of \JWST{} prompts several questions.
What are the critical properties of the high-redshift galaxy population that can be measured with
\JWST{} and why are they interesting?
How well have prior facilities constrained the abundance and properties of high-redshift galaxies?
What do we know about the impact of galaxy formation on the  \reion{} process that
transitioned the neutral intergalactic medium to an ionized state by $z\sim6$? Importantly,
how can \JWST{} improve our knowledge and what new questions can observations with
\JWST{} resolve that have been previously unanswerable to date?

We provide the physical motivations for the study of high-redshift galaxy populations, and
outline what measurements provide critical insight. 
We summarize the progress achieved
in understanding high-redshift galaxies with \HST{} and contemporary observations. 
We review
the capabilities and instrumentation that will allow \JWST{} to make further discoveries,
connecting the capabilities of the instrument to scientific questions they can optimally
address. We conclude by reviewing the observing programs to be conducted during \JWST{} Cycle 1 that will
contribute to answering the outstanding scientific questions our present facilities have
not yet solved. For a detailed summary of the discoveries of \HST{} and \Spitzer{} for
early galaxies, we point the reader to \citet{stark2016a}. For a review of the evolving
properties of the intergalactic medium, see \citet{mcquinn2016a}. A recent relevant
overview of \Lya{} emitting galaxies can be found in \citet{ouchi2020a}.

\section{PHYSICAL COMPONENTS OF COSMIC REIONIZATION}
\label{sec:properties}

\JWST{} can pursue several important goals in characterizing the contribution
of galaxies to \reion{}. These efforts include taking a census of the
ultraviolet radiation supplied by early star-forming galaxies, 
providing indirect
constraints on the Lyman continuum photon production rate of distant galaxies,
contributing to any possible lever on the Lyman continuum escape fraction,
and detecting the possible
presence of evolved stellar populations indicating prior star formation to
very high redshifts. Below, we discuss how \reion{} may be constrained and
review
the current empirical evidence from \HST{} and other facilities for the
contribution of star-forming galaxies.

\subsection{Constraining the Reionization Process}

The term reionization encapsulates the complex process of transitioning
intergalactic hydrogen gas from a mostly neutral to a nearly fully ionized
state. Classically, this process has been represented by the 
average volume-filling
fraction of ionized gas $\Q(z)$ \citep[e.g.,][]{madau1999a}. 
Connecting the ionized fraction $\Q$ with the properties of the earlier galaxy
population amounts to tracking the balance of ionizations and recombinations
in the intergalactic medium, where star-forming galaxies and active
galactic nuclei drive the ionization by supplying Lyman continuum photons.
The process of cosmic reionization involves a complex interplay between the
star formation in galaxies, the character of stellar populations that form,
interior structure of interstellar gas in galaxies, the ionization fraction
of the circumgalactic medium around galaxies, the evolving mean free path of
ionizing radiation in the intergalactic medium, and both the
current strength and history of background radiation from distant sources.
For the purposes of connecting reionization to the galaxy 
population, and using that connection as motivation in part to explore
the status of current high-redshift observations and future promise of
\JWST{} observations, the hydrogen ionization fraction $\Q$ will be treated as a 
global quantity averaged over large scales.
The ionization fraction $\Q$ primarily influenced by a competition between
the production rate of ionizing photons and hydrogen recombinations.

Crudely, for star-forming galaxies this production of ionizing photons can
be captured by the relation
\begin{equation}
\label{eqn:nion}
\nion = \fesc \xiion \rhoUV
\end{equation}
\noindent
where $\rhoUV$ is the comoving total UV luminosity density
($\mathrm{erg}\,\mathrm{s}^{-1}\,\mathrm{Hz}^{-1}\,\mathrm{Mpc}^{-3}$), $\fesc$ is the fraction of 
Lyman continuum photons that escape galaxies
to ionize intergalactic hydrogen, and $\xiion$ is the ionizing photon production efficiency
($\mathrm{Hz}\,\mathrm{erg}^{-1}$)
that indicates the number of Lyman continuum photons per unit UV luminosity density the
stellar populations in galaxies generate. The quantity $\nion$ represents the
comoving density of Lyman continuum photons produced per unit time available for 
ionizing hydrogen in the IGM.

%
%

\subsection{Lyman Continuum Escape Fraction}
\label{sec:lyc}

Constraints on the Lyman continuum 
escape fraction during the reionization era are indirect. The opacity of the
intergalactic medium to ionizing radiation increases dramatically with redshift
\citep{madau1995a,inoue2014a}, effectively preventing the direct detection
of Lyman continuum (\Lyc{}) photons for galaxies above $z\sim4.5$ and making observations
at $z\sim2-4$ difficult. Nonetheless, substantial effort
has been invested in performing direct observations of the Lyman continuum
spectroscopically and photometrically, both in detecting individual sources
and by stacking multiple non-detections.
The vital inferences about Lyman
continuum escape at $z\sim2-4$ enabled by
these efforts can then be applied, with caution, to analogous galaxies
present at $z\gtrsim6$ with the hope of at least motivating assumptions
on $\fesc$ during reionization.

The reliable detection of rest-frame Lyman continuum in distant galaxies
proves challenging owing to the possible foreground contamination by
objects at lower redshifts. The Lyman continuum flux of an object
can be measured via photometry or spectroscopy blueward of $\lambda=912$\AA.
The potential of interlopers means that high-resolution imagery, such as
that provided by \HST{}, is a key component of \Lyc{} direct detection campaigns.
\HST{} provides the ability to resolve nearly coincident interlopers using
non-ionizing flux \citep{mostardi2015a,shapley2016a,pahl2021a} and to assess
directly the \Lyc{} with 
reduced fear of contamination \citep{siana2015a,vasei2016a,rutkowski2017a,fletcher2019a}.
Ground-based imaging in $U$-band or similar filters
remains an attractive approach for constraining
\Lyc{} at high-redshift, given their acceptable
efficiency and availability on large telescopes
\citep{guaita2016a,naidu2018a,mestric2020a}.
Imaging redshifted \Lyc{} also benefits from spectroscopic redshifts to 
prevent non-ionizing photons from a source at an
uncertain distance contaminate the inference of escaping ionizing flux,
or at least quantify the degree of contamination.
Given the large intrinsic ratio of non-ionizing to ionizing flux in the
rest-frame ultraviolet spectra of these galaxies, and the small expected
escape fractions, even a small amount of non-ionizing flux mistaken as
\Lyc{} can render conclusions about $\fesc$ unreliable. Spectroscopic determinations
do not suffer from this uncertainty, but \Lyc{} campaigns with
ground-based slit spectrographs are
in principle not immune to foreground contamination. Perhaps a bigger practical
challenge is conducting large campaigns with blue-sensitive spectrographs, which
involves either the \HST{} Cosmic Origins Spectrograph at low redshift \citep{izotov2016a,izotov2016b,izotov2018a}
or significant
investment with instruments like VLT VIMOS \citep{debarros2016a,marchi2017a,marchi2018a}
or Keck LRIS \citep{shapley2016a,steidel2018a}.
For all high-redshift observations, careful consideration of the effects of
absorption by the intervening IGM is required, including potential variations
in the transmission along different lines of sight \citep[e.g.,][]{shapley2016a,bassett2021a}

Constraints on the escape fraction of \Lyc{} photons have varied widely 
with redshift\footnote{We discuss``absolute'' escape fractions unless otherwise noted.}.
Through \HST{} images of resolved stars in  the
$z\sim0$ galaxy NGC 4214, \citet{choi2020a} determined an
escape fraction of $\fesc\approx0.25$, noting that differences with
previous measurements finding $\fesc\sim0$ could be accounted
for by viewing angle and geometrical effects.
Using \HST{} COS observations of relatively nearby ($z\lesssim0.3$) galaxies,
the \Lyc{} escape fractions measured are reported in the range $\fesc\sim0.06-0.35$ \citep{izotov2016a,izotov2016b,izotov2018a,izotov2021a}.
GALEX-based \Lyc{} constraints provide $\fesc\lesssim0.021-0.095$ at $z\sim1$ \citep{rutkowski2016a}.
Observations with the \HST{} Solar Blind Channel provide limits of $\fesc\lesssim0.04-0.13$
for individual objects at $z\sim1.3$ \citep{alavi2020a}.
Stacks of GALEX non-detections provide limits of $\fesc\lesssim0.028$ at $z\sim2.2$ \citep{matthee2017a},
while non-detections in \HST{} F275W imaging provides $\fesc\lesssim0.056$ at $z\sim2.5$ \citep{rutkowski2017a}.
Using the Hubble Deep UV Legacy Survey, \citet{jones2021a} detected \Lyc{} from five $F275W$ sources
at $2.35<z<3.05$ and concluded galaxies in this redshift range could maintain the IGM ionization state.
\citet{smith2020a} use UV \HST{} imaging in GOODS to measure $\fesc\simeq0.28_{-0.04}^{0.20}$ for AGN
at $z\sim2.3-4.3$.
Through searches of \HST{} UV images of the Ultra Deep Field, \citet{japelj2017a} placed relative escape fraction
limits of $f_\mathrm{esc,rel}<0.07$ for faint galaxies at $z\sim3-4$.
At higher redshift $z\sim3$, more individual detections have been possible \citep[e.g.,][]{mostardi2015a,vanzella2016a}.
The sample-averaged escape fraction from a dozen individual detections 
at $z\sim3$ has been reported as $\fesc=0.06\pm0.01$
from the Keck Lyman Continuum Spectroscopic Survey augmented with \HST{} imaging
\citep[][; updating from $\fesc0.09\pm0.01$ measured by \citealt{steidel2018a}]{pahl2021a}.
At $z\sim3.1$, the LymAn Continuum Escape Survey (LACES) measured \HST{} $F336W$ photometry
for 61 strong line-emitting sources, finding an average escape fraction of $\fesc\sim0.2$
for detected sources and limits in non-detected sources of $\fesc\lesssim0.005$ \citep{fletcher2019a}.
\Lyc{} detection from long-duration gamma-ray burst host galaxies at $z\sim3-3.5$
implies escape fractions of $\fesc\approx0.08-0.35$ \citep{vielfaure2020a}.
A remarkable individual \Lyc{} detection at $z=4$ with VLT FORS suggests a relative escape fraction
of $f_{\mathrm{esc,rel}}=0.5-1.0$ \citep{vanzella2018a}.

Given the range of measurements for $\fesc$ over a wide range of redshifts $z\lesssim4$, definitive conclusions
about \Lyc{} escape during \reion{} at $z\gtrsim6$ are difficult to draw. The direct measurement
of \Lyc{} is critical in establishing the reality of ionizing flux escaping from galaxies, but
the process of \Lyc{} and its connection to other observables needs to be understood to 
extrapolate measurements at $z\lesssim4$ to the era of \reion{}.
Other methods for estimating $\fesc$ for more distant objects are needed. 
Cross-correlations between galaxies and IGM transmissivity in sightlines to distant
quasars have enabled escape fraction of $\fesc=0.23_{-0.12}^{0.46}$ for star-forming galaxies at $z\sim5.5-6.4$ to be inferred 
\citep[][]{kakiichi2018a,meyer2020a}.
Experiments measuring redshifted 21cm signals will provide an
indirect handle on $\fesc$ via constraining the impact of galaxies
on the evolving
properties of the IGM \citep[e.g.,][]{hera2021a,hera2021b}.
Consistency checks between the strength of observed nebular emission from high-redshift
galaxies, which is powered by the ionization of the interstellar medium by
Lyman continuum emission internal to galaxies, the escape fraction, and the
character of the stellar population expressed by $\xiion$ can be performed. However,
without direct Lyman continuum detections the
model-independent inference of $\fesc$ will likely remain elusive for high-redshift galaxies.

The predicted extreme absorption by the IGM suggests a high
improbability of reliable direct detections of \Lyc{} emission
during \reion{}. To connect direct \Lyc{} observations at lower
redshifts with the reionization process, the physics of
\Lyc{} production and escape need to be understood. As the
physical process of \Lyc{} escape becomes clearer, observed
correlations between the inaccessible $\fesc$, the production efficiency
of ionizing photons $\xiion$, and observationally accessible
signatures can be used to interpret more confidently how
leaking \Lyc{} from high-redshift galaxies reionizes intergalactic
hydrogen.

\subsection{Physics of Lyman Continuum Escape}

For understanding the physics of \Lyc{} escape, astrophysical
simulations play a critical role. Large-scale simulations
that predict how the properties of the galaxy population
evolve during \reion{} are presented in Section \ref{sec:models}.
Here, we highlight a few detailed simulations of 
\Lyc{} escape in galaxies directly relevant for understanding
how $\fesc$ is determined rather than studying its implications.
Using radiative transfer applied to zoom-in simulations of
galaxy formation coupled with
synthesis models for predicting the emission from stellar
populations, \citet{ma2016a} and \citet{ma2020a} found that
feedback from star formation supports an escape fraction
of $\fesc\approx0.2$. In high mass galaxies, dust reduces
$\fesc$, while at the extreme low-mass end $\fesc$ again
decreases owing to the inefficiency of feedback in creating
escape paths for ionizing radiation. Using radiative
transfer post-processing of zoom-in cosmological simulations
including a detailed model for feedback, \citet{barrow2020a}
found that when \Lyc{} escapes from star-forming galaxies
it does so through low column sightlines affected by 
prior feedback from star formation. Computing possible
observable signatures associated with high $\fesc$,
they found that while high $\fesc$ occurs during
periods of strong [OIII]/[OII] line ratios associated with
ionizing input from massive stars, the converse was not
necessarily true and that no direct causal connection existed
between strong [OIII]/[OII] and $\fesc$
\citep[see also][for a similar observational inference]{izotov2017a}. 
In radiation hydrodynamical simulations with $<5$pc resolution,
\citet{secunda2020a} found that since 
periods of high $\fesc$ follow vigorous star formation episodes
by $\sim100$Myr, the delayed increase in ionizing photons
associated with binary evolution after the initial stellar formation
leads to an increased photon-weighted mean escape fraction of
$\fesc\approx0.17$.
Zoom-in cosmological simulations including radiative hydrodynamics
can be used to predict the connection between $\fesc$ and
near UV lines, and CII $1334$\AA{} flux correlates well with $\fesc$
after correcting for dust attenuation \citep{mauerhofer2021a}.
Using semi-analytical models, \citet{seiler2018a} found
enhanced $\fesc$ for intermediate mass galaxies where phases
of AGN feedback could affect the subsequent ability of ionizing
photons to escape. On much smaller scales, using radiation
hydrodynamical simulations of \Lyc{} escape from HII regions,
\citet{kakiichi2021a} found that \Lyc{} photons leak through
low column density regions in the turbulent interstellar
medium. Supersonic turbulence in the ISM associated with
star-forming clouds leads to a wide density and surface
density distribution, providing low column pathways for
ionizing photons to escape even before supernovae feedback.

\subsection{Indirect Measures of Lyman Continuum Escape}

Constructing observational measures of \Lyc{} escape
that do not rely on direct detection of \Lyc{} emission
are essential for understanding the reionization era.
These techniques will prove increasingly important
during \JWST{} operations as rest-frame optical spectra
will become a standard tool during the study of galaxies
during \reion{}. Related indirect measures
of the ionizing photon efficiency $\xiion$ are discussed
in Sections \ref{sec:xiion} and \ref{sec:spec_astro}.
Here, we review studies of
indirect measures of $\fesc$ in galaxies. 

Local observations have provided interesting test beds
for studying the connection between $\fesc$ and 
other galaxy properties. \citet{alexandroff2015a}
found that for a sample of far-UV COS-observed galaxies,
\Lyc{} leakiness most strongly correlated with
\Lya{} equivalent width.
Local galaxy observations with COS suggest that
\Lyc{} $\fesc$, the escape of \Lya{}, and the \Lya{}
equivalent width are connected \citep{izotov2018a,izotov2020a}, but
find no clear connection between high [OIII]/[OII] and high $\fesc$. 
Escape fractions in these objects
decrease with increasing velocity separation between
the peaks of \Lya{} emission, and this relationship has been
used to estimate an escape fraction of $\fesc\approx0.59$
in the lensed $z=6.803$ galaxy A370p-z1 \citep{meyer2021a}.
Regarding the possible role of outflows in increasing $\fesc$,
\citet{chisholm2017a} found that \Lyc{} leaking galaxies fell on the
low side of the equivalent width distribution of absorption lines
tracing outflows, suggesting low metallicities and low HI column densities.
Using UV Lyman series lines and low-ionization metal absorption
lines, \citet{chisholm2018a} found that HI and ISM absorption
lines could be used to predict $\fesc$ in low-redshift \Lyc{} leakers.
Recently, using a sample of more than a dozen confirmed \Lyc
emitting, mostly low-redshift galaxies \citep{gazagnes2020a} determined  
correlations between low neutral gas covering fraction $f_{\mathrm{cov}}$, low \Lya{} peak velocity
separations, and high $\fesc$.

At higher redshifts, both optical emission and ISM absorption
have been connected to $\fesc$.
Using almost a thousand $z\sim3$ galaxies, \citet{reddy2016a}
established an empirical correlation between the reddening $E(B-V)$ and
neutral gas covering fraction $f_\mathrm{cov}$. If the absolute
\Lyc{} escape fraction is $\fesc\approx 1-f_\mathrm{cov}$, then
reddening and $\fesc$ can be related and estimated in distant
galaxy populations.
In their spectral analysis at $z\sim3$, 
\citet{steidel2018a} found a strong correlation between
$\fesc$ and \Lya{} equivalent width, which induced
a residual correlation between $\fesc$ and UV luminosity.
Using rest-frame optical spectra of the $z\sim3$ \Lyc{}
leaking galaxies from LACES, \citet{nakajima2020a} 
demonstrated an empirical connection between large
[OIII/OII] emission line ratios and large $\fesc$
in their sample.
In principle, any of these methods could be honed for
applications to high-redshift galaxy populations, especially
as multiplexed IR spectroscopy from \JWST{} is a near-term possibility.


\subsection{Ionizing Photon Production Efficiency}
\label{sec:xiion}

The quantity $\xiion$ provides a translation between the observable
rest-frame UV emission from galaxies and the corresponding number
of Lyman continuum photons their stars produce. If the Lyman continuum
emission from galaxies was directly observable in the reionization era,
the product $\fesc\xiion L$ from a single galaxy could be replaced
by the emergent Lyman continuum luminosity $\Lion$ normalized
by average photon energy to get the number of Lyman continuum
photons each galaxy releases. Instead, the
proxy $\xiion$ is used to reflect how the observed 
UV spectrum of the galaxy is expected to extend blueward
of $\lambda=912$\AA. The expectations for $\xiion$ are
model-dependent, as they rely on the initial mass function
of stars, the galaxy's star formation history,
the evolution of individual stars, stellar metallicity
that affects their mass-loss rates and sizes, and potential
stellar binary interactions \cite[e.g.,][]{zackrisson2011a,zackrisson2013a,zackrisson2017a,eldridge2017a,stanway2018a,stanway2019a}.
In what follows, we use the Binary Population and Spectral 
Synthesis model \citep[hereafter BPASS;][v2.2 used throughout]{eldridge2017a} to explicate the connection between metallicity,
binarity, and the ionizing photon production efficiency $\xiion$.

Metallicity influences ionizing photon production efficiency
in several ways. The amount of mass loss to stellar winds depends on 
the stellar metallicity by changing the atmospheric opacity. 
Stars that lose mass through winds also lose angular momentum, 
which in turn reduces the replenishment of fuel through
rotational mixing. The high-metallicity
stars then burn slower at lower temperatures over longer timescales
than low-metallicity stars. Metallicity also has a strong effect
on the importance of binary interactions.
As discussed in \citet{eldridge2017a}, binary star interactions 
lead to substantial mass transfer from the primary to the
secondary star. The accretion of mass onto the secondary
can lead to an increased rotation rate and, if more than
$\sim5\%$ of the initial stellar mass is accreted, the rotation
will lead to rotational mixing of fuel inside the star and
more vigorous burning. The
result is a substantial increase in the ionizing photon
production rate relative to singleton stars. High metallicity
stars lose more mass to stellar winds, which leads to more
compact older stars and less mass transfer in binary interactions.
By combining a detailed description of these physics with a
model for the fraction of stars that occur in binaries and 
the distribution of binary orbits, the resulting enhancement
to the overall ionizing
photon production efficiency of the population owing jointly to 
metallicity and binary interactions can be computed.

\begin{figure}[h]
\includegraphics[width=4in]{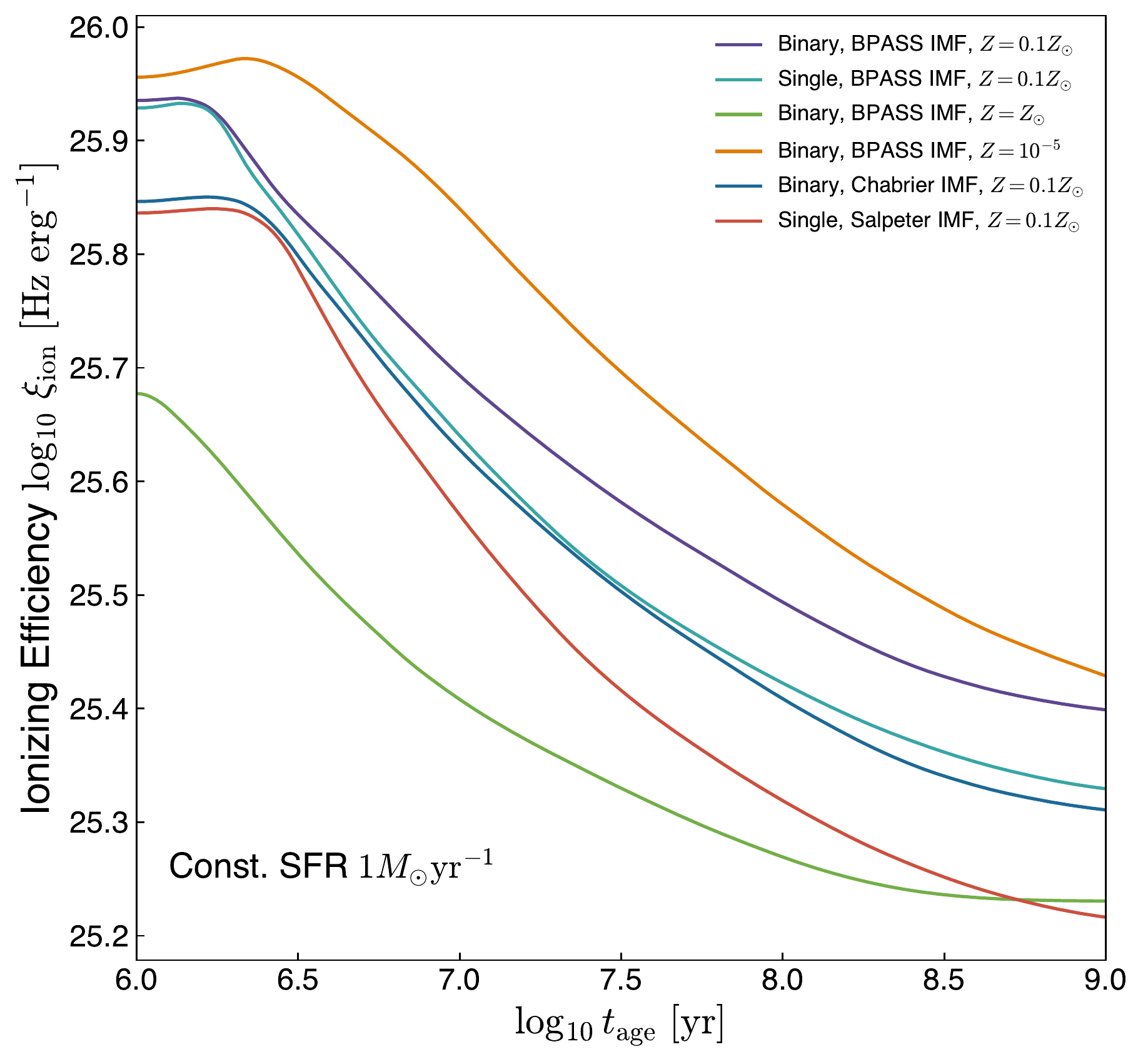}
\caption{Ionizing photon production efficiency $\xiion$ as a 
function of stellar population age in the BPASS population synthesis
models \citep{eldridge2017a,stanway2018a,stanway2019a}. Shown
is the ratio $\xiion\equiv\dot{N}_{\mathrm{ion}}/L_{1500\mAA}$
with units $[\mathrm{Hz}\,\,\mathrm{erg}^{-1}]$,
where $\dot{N}_{\mathrm{ion}}$ and $L_{1500\mAA}$ are respectively
the 
number of ionizing photons per second and the specific UV luminosity
near 1500\AA{} produced by a constant star formation rate population
with age $t_{\mathrm{age}}$. The model predictions cover both
single and binary star evolution that affect the predicted
temperature of massive stars, as well as metallicity effects
that cause low metallicity stars to prove more efficient at
producing ionizing radiation than higher metallicity stars.
The ionizing photon production is also influenced by the initial
mass functions (IMFs) with the default BPASS \citep[see text and][]{eldridge2017a}, \citet{chabrier2003a}, and \citet{salpeter1955a} IMF model predictions shown for $Z=0.1\Zsun$
metallicity. Note that these curve labels assume $\Zsun\equiv0.02$.}
\label{fig:xi_ion}
\end{figure}

Figure \ref{fig:xi_ion} shows the evolution of $\xiion$ predicted
by BPASS for a dust-free galaxy with a  
constant star formation rate as a function of stellar
population age $\tage$. Shown
are models with (``Binary'') and without (``Single'') binary
interactions, with different metallicities $Z$ relative to the
solar value (assumed to here be $\Zsun=0.02$). The production
of ionizing photons depends on the presence of massive stars
and therefore the initial mass function (IMF). In BPASS, the
initial mass function is modeled as a broken power-law
such that the number of stars with masses less than a maximum
mass $\mmax$ is given by
\begin{equation}
\label{eqn:imf}
N(m<\mmax) \propto \int_{0.1}^{m_1}\left(\frac{m}{\Msun}\right)^{\alpha_1}
dm + m_1^{\alpha_1} \int_{m_1}^{\mmax} \left(\frac{m}{\Msun}\right)^{\alpha_2} dm
\end{equation}
where $m_1$ is the mass where the power-law breaks and $\alpha_1$
and $\alpha_2$ are the power-law slopes for low mass and high mass
stars, respectively. 
The default BPASS IMF has 
$\alpha_1=-1.3$, $\alpha_2=-2.35$, $m_1=0.5\Msun$ and $\mmax=300\Msun$.
In this formulation, the \citet{salpeter1955a}
IMF is a continuous power-law with $\alpha_1=\alpha_2 = -2.35$.
The \citet{chabrier2003a} has a modified exponential cut-off at low
masses ($m_1=1\Msun$), and a slightly shallower high mass slope ($\alpha=2.3$). The models in Figure \ref{fig:xi_ion} assume 
a maximum mass $\mmax=100\Msun$ for the \citet{salpeter1955a} and
\citet{chabrier2003a} IMF models. 

The differences in $\xiion$ with binarity, metallicity, IMF, and 
population are substantial \citep{stanway2016a}. For binary models, low metallicity
populations produce $2-3\times$ more ionizing photons per unit
UV luminosity than solar metallicity populations. At fixed
metallicity, binary interactions increase $\xiion$ by about
$25\%$ at $\tage=100$Myr. At fixed metallicity, the default
BPASS IMF results in a higher $\xiion$ than \citet{chabrier2003a}
also by $\approx25\%$. 
For reference, a stellar population of age $\tage\sim100$Myr and
metallicity of $Z=0.1\Zsun$ will have
$\log_{10} (\xiion/\mathrm{Hz}\,\mathrm{erg}^{-1}) \approx 25.46$.
For a metallicity of $Z=0.005\Zsun$,
$\log_{10} (\xiion/\mathrm{Hz}\,\mathrm{erg}^{-1}) \approx 25.60$.
Direct metallicity constraints from ALMA OIII $88\um$ at $z>7$
indicate $Z\approx0.2\Zsun$ \citep{jones2020a}, such that large $\xiion$ should be possible. 
The $\xiion$ of the entire population will depend on the
collection of star formation histories from individual objects.
For instance, in the BLUETIDES simulations \citet{wilkins2016a}
found variations in the star formation histories and metal
enrichment in high-redshift galaxies led to 
$\log_{10} (\xiion/\mathrm{Hz}\,\mathrm{erg}^{-1}) \approx 25.1-25.5$.

From the strength of nebular emission powered by reprocessed
\Lyc, observations can provide estimates of $\xiion$.
At $z\sim2$, estimates of $\xiion$ from
UV complete samples with H$\alpha$ spectroscopy give
$\log_{10} (\xiion/\mathrm{Hz}\,\mathrm{erg}^{-1}) \approx 25.47$
\citep{emami2020a}.
MOSDEF has constrained ionizing photon production
spectroscopically at $z\sim2$,
finding a lower limit of $\log_{10} (\xiion/\mathrm{Hz}\,\mathrm{erg}^{-1}) \approx 25.06-25.34$
depending on the dust model, with elevated $\xiion$ in galaxies with high [OIII]/H$\beta$ or with
UV continuum slopes $\beta<-2.1$ \citep{shivaei2018a}.
Star-forming galaxies at redshifts $z\sim2-3$ often show increased \Lya{} 
equivalent widths with increasing
[OIII]/H$\beta$, indicating a connection between enhanced ionizing photon production 
efficiency and escape \citep{tang2020a}.
At redshift $z\sim3.1-3.7$, \citet{nakajima2016a} and \citet{onodera2020a} found evidence for ionizing photon
production efficiencies as high as $\log_{10} (\xiion/\mathrm{Hz}\,\mathrm{erg}^{-1}) \approx 25.8$
in strong line emitting galaxies.
Observational estimates of $\xiion$ at $z\sim4-5$, where H$\alpha$
nebular emission is still probed by \Spitzer{} broadband colors,
provide values of
$\log_{10} (\xiion/\mathrm{Hz}\,\mathrm{erg}^{-1}) \approx 25.36-25.8$
\citep{bouwens2016a,lam2019a}, with strongest $\xiion$ in objects
with blue UV continua ($\beta<-2.3$). 
The same approach can be applied out to very high redshift using the
enhanced \Spitzer{} coverage of GOODS from the GREATS survey, and
\citet{debarros2019a} measured $\log_{10} (\xiion/\mathrm{Hz}\,\mathrm{erg}^{-1}) \approx 25.77$ for early galaxies at $z\sim8$.
Possible AGN contribution to these ionizing photon production efficiencies can in principle
be determined by measuring rest-UV spectral lines, comparing the collisionally excited
metal lines with the He $1640$\AA{} recombination line, or by using UV Ne line diagnostics \citep{feltre2016a}.

\subsection{Luminosity Density}
\label{sec:rhoUV}

The total UV luminosity density $\rhoUV$ provides an observable handle on the
cosmic star formation rate density $\rhoSFR$, as the rest-frame UV emission
from a galaxy is connected to the presence of newly-formed massive stars
\citep[see the review by][]{madau2014a}.
The luminosity density $\rhoUV$ depends on the integral
of the luminosity function $\Phi(L)\equiv dn/dL$ weighted by the UV luminosity $L$ 
(a luminosity density in the rest-frame UV, typically taken near $\lambda=1500$\AA,
with units $\mathrm{erg}\,\mathrm{s}^{-1}\,\mathrm{Hz}^{-1}$) as
\begin{equation}
\rhoUV = \int_{0}^{\infty} \Phi(L) L dL 
\end{equation}
\noindent
where the formal limits $[0,\infty]$ are effectively truncated at low and high luminosities
by the physics of galaxy formation that determine the functional form of $\Phi(L)$.
Indeed $\rhoUV$ is finite, even as the observed shape of $\Phi$, typically close to a 
\cite{schechter1976a} function
\begin{equation}
\label{eqn:schechterL}
\Phi(L) = \phi_{\star}\left(\frac{L}{\Lstar}\right)^{\alpha}\exp\left(-\frac{L}{\Lstar}\right) d\left(\frac{L}{\Lstar}\right)
\end{equation}
\noindent
can approach a divergent faint-end slope $\alpha=-2$ over the range
of observed luminosities. In terms of the UV absolute magnitude $\MUV$,
the \cite{schechter1976a} function can be written as
\begin{eqnarray}
\label{eqn:schechter}
x&\equiv&10^{-0.4(\MUV - \Mstar)},\\
\phi(\MUV)&=& 0.4\ln 10\,\phistar x^{(\alpha+1)} \exp\left(-x\right),
\end{eqnarray}
\noindent
\noindent
where typically we find
the normalization $\phistar\lesssim10^{-3}$ [$\mathrm{Mpc}^{-3}\,\mathrm{mag}^{-1}$] 
and the characteristic magnitude $\Mstar\approx-21$ at $z\gtrsim6$.
The functional form Equations \ref{eqn:schechterL} and \ref{eqn:schechter} should
be considered approximate, as eventually at low enough luminosities the 
efficiency of galaxy formation should truncate the abundance of faint systems
through a limit on cooling processes, molecular gas formation, or feedback \cite[e.g.][]{sun2016a}.
Additionally, given a large enough survey area to discover rare, bright galaxies, the abundance
of objects at $\MUV<-22$ can be determined. These luminous galaxies at $z\gtrsim6$ occur
in an abundance larger than \citet{schechter1976a} function fits predict, given the
exponential cut-off at bright magnitudes. Instead, a double power-law (DPL) fit can be used
with a functional form given by
\begin{eqnarray}
\label{eqn:dpl}
y&\equiv&10^{-0.4(\MUV - \Mstar_{\mathrm{DPL}})},\\
\phi_{\mathrm{DPL}}(\MUV) &=& 0.4\ln 10\,\phistar_{\mathrm{DPL}} \left[y^{-(\alpha+1)} + y^{-(\beta+1)}\right]^{-1}.
\end{eqnarray}
\noindent
Here, the parameter $\beta$ describes the power-law behavior brighter than the
characteristic magnitude $\Mstar_{\mathrm{DPL}}$. Fit to the same data,
Equations \ref{eqn:schechter} and \ref{eqn:dpl} are not guaranteed to 
provide the same normalization or characteristic magnitude, and we will
denote the DPL parameters as $\phistar_{\mathrm{DPL}}$ and $\Mstar_{\mathrm{DPL}}$
to distinguish them.

\begin{figure}[h]
\includegraphics[width=\linewidth]{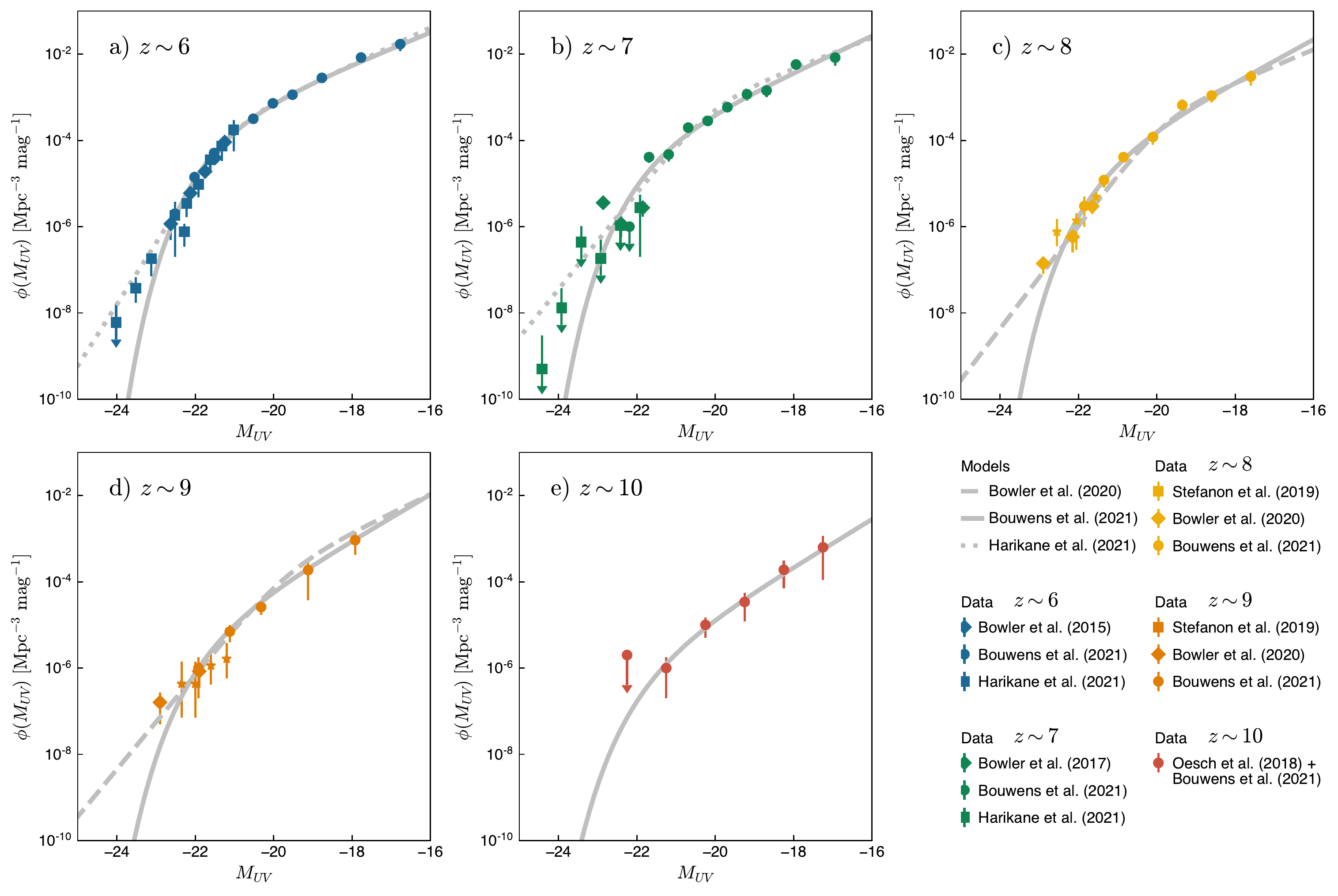}
\caption{Rest-frame ultraviolet luminosity function of galaxies
during \reion{} at redshifts $z\approx6$ (blue), $7$ (green), $8$
(yellow), $9$ (orange), and $10$ (red).
Shown are data compiled from the most recent available
literature at each magnitude, using both ground-based and \HST{} observations to identify high-redshift
 galaxies based primarily on dropout color selections \citep{bowler2015a,bowler2017a,oesch2018a,stefanon2019a,bowler2020a,bouwens2021a,harikane2021a}. Also shown are \citet{schechter1976a} function 
\citep[solid lines][]{bouwens2021a} and double power-law
\citep[dotted and dashed lines]{bowler2020a,harikane2021a} 
model fits that provide representations of the data with
comparable quality.
Empirically, the abundance of galaxies
declines from $z\sim6-10$, with
the faint-end slope steepening to $\alpha<-2$ to
roughly track the low-mass slope of the halo mass function. The
Schechter function fits have characteristic magnitudes
$M_\star\approx -21$ at all redshifts shown, while
the pivot magnitude of the double power-law fits 
dims from $M_\star^{\mathrm{pl}}\approx-21$ at $z\sim6$
to $M_\star^{\mathrm{pl}}\approx-19.7$ at $z\sim9$.}
\label{fig:lf}
\end{figure}

Figure \ref{fig:lf} shows determinations of the rest-frame UV luminosity function at 
redshifts $z\sim6$, $7$, $8$, $9$, and $10$ collected from a small subset
of the literature. 
The search for high-redshift galaxy populations and
quantifying their $\rhoUV$ evolution
have been extensive undertakings by the community
\citep[e.g.,][]{conselice2016a,mcleod2016a,finkelstein2021a}. The
data in Figure \ref{fig:lf} are by no means complete.
The 
measurements in Figure \ref{fig:lf} 
include galaxies identified in both ground-based and \HST{} imaging, using 
Lyman break drop-out selection techniques. The bright ends of the luminosity
functions come from \citet{harikane2021a} at $z\sim6-7$ and \citet{bowler2015a},
\citet{bowler2017a}, \citet{stefanon2019a}, and \citet{bowler2020a} at 
redshifts $z\sim8-9$. At redshifts $z\sim6-9$, most data fainter than $\Mstar$
come from \citet{bouwens2021a}. At $z\sim10$, the measurements by \citet{oesch2018a}
are used, as updated in \citet{bouwens2021a}. Also shown are representative
Schechter luminosity function models at $z\sim6-9$ from \citet{bouwens2021a}
and $z\sim10$ from \citet{oesch2018a},
and DPL models from \citet{harikane2021a} at $z\sim6-7$ and from \citet{bowler2020a}
at $z\sim8-9$. The luminosity function
model parameter values are reproduced in Tables \ref{tab:schechter} and \ref{tab:dpl}.

The redshift-dependent luminosity functions provide an interesting picture for the 
evolution of galaxies in the \reion{} era. The overall abundance of galaxies
brighter than $\MUV\approx-16$ declines strongly from $z\sim6$ to $z\sim10$.
The luminosity function faint-end slope is steep, with $\alpha\lesssim-2$, but
the degree of steepening with redshift depends on whether a Schechter or DPL
model is used.
The luminosity density $\rhoUV$ (ergs s$^{-1}$ Hz$^{-1}$ Mpc$^{-3}$)
can be computed by integrating the luminosity function weighted by
galaxy luminosity. Table \ref{tab:rhoUV} reports the luminosity
density contributed by galaxies with luminosities $L>0.01L^{\star}$,
for the luminosity function models shown in Figure \ref{fig:lf}
with best-fit parameters provided in Tables \ref{tab:schechter} and \ref{tab:dpl}.
Put in the cosmological context, over the range of redshift $z\sim8-10$
the luminosity density in the universe changes by roughly an 
order of magnitude in less than 200 Myr.
Table \ref{tab:rhoUV} also reports
the fractions of the luminosity densities at each
redshift contributed by galaxies with luminosities $L>0.1L^{\star}$
and $L>L^{\star}$. 
The inferred relative importance of bright versus faint
objects also depends on the luminosity function model used, reflecting
the ability of the DPL model to capture the presence of extremely
bright objects above the Schechter function exponential cut-off.
Given the sharp drop of the Schechter model, even a small number of
bright objects can provide number densities well in excess of the
Schechter curve at $M<\Mstar$. The total luminosity densities
between the Schechter and DPL model fits differ by factors of
$\approx1.05-2$, influenced by the differing contributions of
bright galaxies. Galaxies brighter than $\Lstar$ can provide
roughly twice as large a relative contribution to $\rhoUV$
in the DPL models compared with the Schechter fits. How these
contributions translate to the relative importance of bright
versus faint galaxies to \reion{} depends on the prior
cosmic star formation history, the variations in $\xiion$
with galaxy luminosity, and how the escape fraction $\fesc$
may depend on other galaxy properties \citep[e.g.,][]{duncan2015a,sharma2016a,lewis2020a,naidu2020a,yung2020a,yung2020b}.
Indeed, in revisiting the relative contribution of 
bright galaxies to reionization \citet{naidu2020a} found that
as much as $\approx45\%$ of the UV luminosity density at
$z\sim7$ could be contributed by objects with $\MUV<-20$.
Understanding the detailed process of \reion{}, including
the expected redshift-dependent topology of ionized hydrogen,
will require further constraints on the relative contribution of
bright and faint galaxies over cosmic time. Forthcoming surveys
with \JWST{}, such as the wide-area COSMOS-Webb program
(see Section \ref{sec:jwst}), should
help resolve this question and could rebalance our views
on the reionizing role of galaxies as a function of luminosity.

In noting the strong evolution in the luminosity density
with redshift during \reion{}, the sense of whether
$\rhoUV$ is changing rapidly relative to expectations
should be calibrated by theoretical models. By
converting the observed
$\rhoUV$ to a star formation rate density
$\dot{\rho}_{\star}$ using stellar population 
synthesis modeling, a comparison with predictions
for the star formation in growing dark matter
halos can be performed. For reference, the
BPASS population synthesis models with binary
stellar evolution \citep{eldridge2017a,stanway2018a,stanway2019a}
predict that a galaxy forming
stars at a constant rate $\psi=1$ Msun yr$^{-1}$
for 100 Myr will have a specific UV luminosity 
of $\log_{10}\LUV\approx28.1$ ergs s$^{-1}$ Hz$^{-1}$.
For a dust-free system this model then suggests an approximate
conversion $\log_{10} \dot{\rho}_{\star}\approx\log_{10} \rhoUV - 28.1$,
which would yield $\log_{10} \dot{\rho}_{\star}(z\sim10)\approx -3.3$
[$\Msun$ yr$^{-1}$ Mpc$^{-3}$].

\begin{figure}[h]
\includegraphics[width=\linewidth]{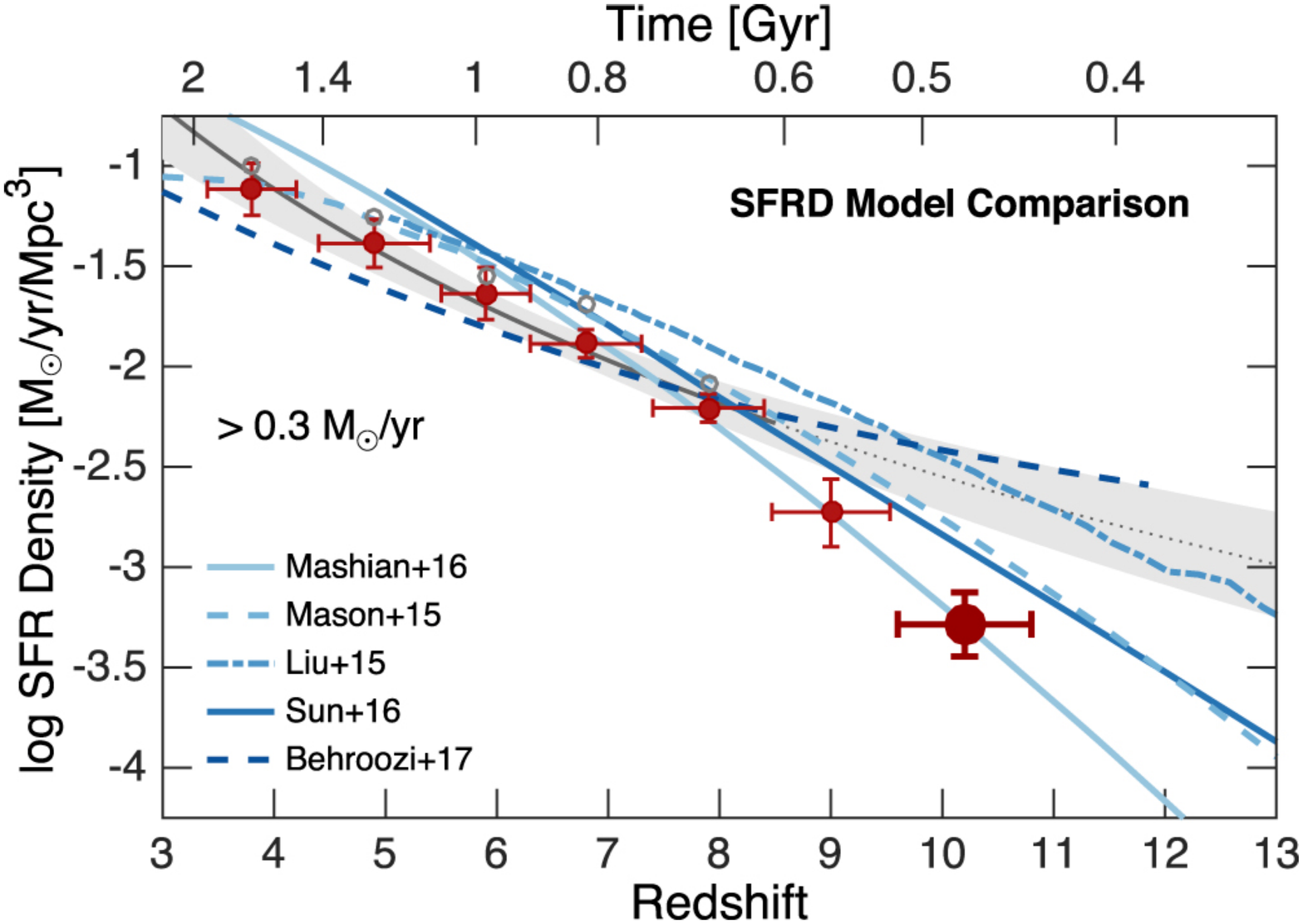}
\caption{Evolution of the star formation rate (SFR) density
inferred from the evolving UV luminosity function.
Shown are the SFR densities converted from integrating the
UV luminosity function over a range of luminosities 
corresponding to a fixed minimum SFR $>0.3\Msun~\mathrm{yr}^{-1}$
(red points with error bars).
Empirically,
the SFR density above this threshold evolves rapidly above
$z\sim8$. Also shown are some theoretical models for the
SFR density evolution \citep[blue lines]{mason2015a,mashian2016a,sun2016a}
and a power-law extrapolation fit to the $z\leq8$
data (gray area). Models with approximately redshift-independent star
formation efficiency with halo mass roughly track the redshift decline seen
in the data \citep[e.g.,][]{mashian2016a}
Figure reproduced from \citet{oesch2018a} with permission.}
\label{fig:oesch}
\end{figure}

Figure \ref{fig:oesch} shows the evolving star 
formation rate determined by \citet{oesch2018a} at
redshifts $z\sim4-10$ from rest-frame UV-selected
galaxies, along with theoretical models for 
$\dot{\rho}_{\star}(z)$ \citep{mason2015a,mashian2016a,sun2016a,liu2016a}.
The best agreement between the theoretical models and
the observed cosmic SFR history occurs for models
where the star formation efficiency as a function of
halo properties remains roughly constant with redshift.
In these models the star formation rate density does
decline rapidly with redshift in step with the abundance of dark
matter halos, although the details of galaxy formation physics
in the smallest halos can affect the normalization of the
trends. While not yet conclusive,
a relatively sharp decline in both $\rhoUV$ and
$\rhoSFR$ above $z\sim8$ may therefore not be unexpected
and may simply tracethe evolution in halo abundance.
Subsequent theoretical models of the high-redshift galaxy
population reach similar conclusions about the connection 
between the redshift-constancy of the star formation efficiency 
as a function of halo mass and the rapid decline in the
star formation rate density \citep{mitra2015a,yue2016a,lapi2017a,yung2019a,yung2019b,yung2020a,yung2020b,behroozi2020a}, as do recent observational analyses
\citep[e.g.,][]{bouwens2021a,harikane2021a}.

\subsection{Evolution of the Ionized Fraction}
\label{sec:Q}

The volume fraction $\Q(z)$ of ionized hydrogen in the IGM provides
a convenient global summary of reionization. Below we describe
models for how $\Q(z)$ evolves, which is subject to significant
uncertainties beyond those associated with the production of
IGM-ionizing radiation from galaxy populations already
described in Sections \ref{sec:xiion}
and \ref{sec:rhoUV}.
The detailed
time
dependence of the ionized fraction has not yet been measured.
The
Lyman-$\alpha$ optical depth measured in the spectra of quasars suggests
reionization must complete (e.g., $\Q>0.9$) by $z\sim5.5-6$ \citep{fan2006a,mcgreer2015a,boera2019a}.
Other constraints have been determined, primarily through interpreting
observations of Lyman-$\alpha$ emission or absorption
at a variety of redshifts $z\sim6-8$, finding
$\Q\sim0.2-0.8$ \citep{davies2018a,mason2018a,mason2019a,hoag2019a,wang2020a,yang2020a,jung2020a}.
Very recently, limits on the 21cm power spectrum at $z\sim8$
and $z\sim10.4$ have been placed by the 
Hydrogen Epoch of Reionization Array \citep[HERA;]{hera2021a}.
These limits can be combined with interpretative simulations 
to constrain the evolution of $\Q(z)$ \citep[][see also \citealt{monsalve2017a} for constraints from the EDGES experiment]{hera2021b}.
There are integral
constraints on $\Q$ from the electron scattering optical depth $\tau$
to the cosmic microwave background (CMB).

\citet{madau1999a} provided a basic picture for the evolution of $\Q(z)$
in the form of a differential equation
for the time rate of change $d\Q/dt$ arising from the
competition between ionization and recombination. In its
simplest form, the
``reionization equation'' can be written
\begin{equation}
\label{eqn:reion}
\frac{d\Q}{dt} = \frac{\nion}{\langle n_H \rangle} - \frac{Q}{\trec},
\end{equation}
\noindent
where $\langle n_H \rangle$ is the mean number density of
hydrogen. The recombination time $\trec$ is intended to
capture possible variations in the recombination rate owing
to density inhomogeneities, which is often represented \citep[e.g.,][]{madau1999a} as
\begin{equation}
\trec = \left[(1+Y/4X)*C\langle n_H \rangle\alpha(T)\right]^{-1}
\end{equation}
where $Y$ is the helium abundance, $X$ the hydrogen abundance,
$\alpha$ the temperature-dependent
recombination coefficient of hydrogen (Case A, often taken at $T=20,000$K),
and the clumping factor $C\sim\langle n_H^2 \rangle/\langle n_H\rangle^2$.
Clumping factors of order $C\lesssim3$ at $z\gtrsim6$ are typically
adopted, motivated by simulations
\citep[e.g.,][]{finlator2012a,shull2012a,robertson2015a,gorce2018a}.
Given the cosmological evolution of the mean density,
the evolution of $\Q$ is monotonic if $\nion$ is monotonic. The
observed evolution in $\rhoUV$ (Section \ref{sec:rhoUV}) makes this
a reasonable assumption, although double reionization models are still
being explored \citep{salvador-sole2017a}. Once $\Q$ is determined,
the neutral fraction can be determined as $\xHI=1-\Q$.

The simplified description of the evolution of the ionized fraction
computed via Equation \ref{eqn:reion}
is still in wide use for interpreting the impact of galaxy evolution
on \reion{} \citep{finkelstein2019a,naidu2020a}. Given the complexity
of the reionization process, the inadequacies of this model should
not be surprising. \citet{madau2017a} recently presented an
augmentation of the reionization equation that accounts for the
presence of dense, neutral absorbers after the IGM becomes mostly 
ionized through an additional opacity term, preventing the
unphysical values of $\Q>0$. Of course, full cosmological
hydrodynamical simulations can be used to model the reionization,
some of which include radiative transfer
(see Section \ref{sec:models}). In cosmological simulations, the
photoionization rate is usually connected to a model for the
evolving emissivity of galaxies (e.g., the $\rhoUV$ in Table \ref{tab:rhoUV})
and time-dependent mean free path of ionizing photons.
The mean free path evolution may be assumed or computed directly
via radiative transfer.
However, most cosmological simulations
adopt a fixed photoionization rate history or adopt a single
model for the ionizing input from stellar populations.
While
Equation \ref{eqn:reion} allows for the convenient computation 
of $\xHI$ given a model for $\nion$, it does not track the physical
impact of photoionization on the intergalactic gas including the
photoheating associated with hydrogen and helium ionizations. The
subsequent thermal and ionized fraction evolution of the IGM has
a host of additional observational ramifications beyond the scope of
this work (see the review by \citealt{mcquinn2016a}), so, in principle,
the evolution of the neutral fraction during reionization and the
properties of the IGM after reionization, such as the \Lya-forest
forest power spectrum, should be made consistent. \citet{villasenor2021a}
used hundreds of hydrodynamical cosmological simulations, varying
the amplitude and timing of the \citet{puchwein2019a} photoionization
and photoheating rates, and determined a photoionization and
photoheating history that matches all the available data on the
\Lya-forest at $z<5$. These fits then provide a prediction for
the neutral fraction evolution $\xHI(z)$ fully consistent with
the post-reionization properties of the IGM. Note this model, like
\citet{puchwein2019a}, assumes $\fesc\approx0.18$ during reionization
but the amplitude and timing of the photoionizing background are allowed to vary.

\begin{figure}[h]
\includegraphics[width=4in]{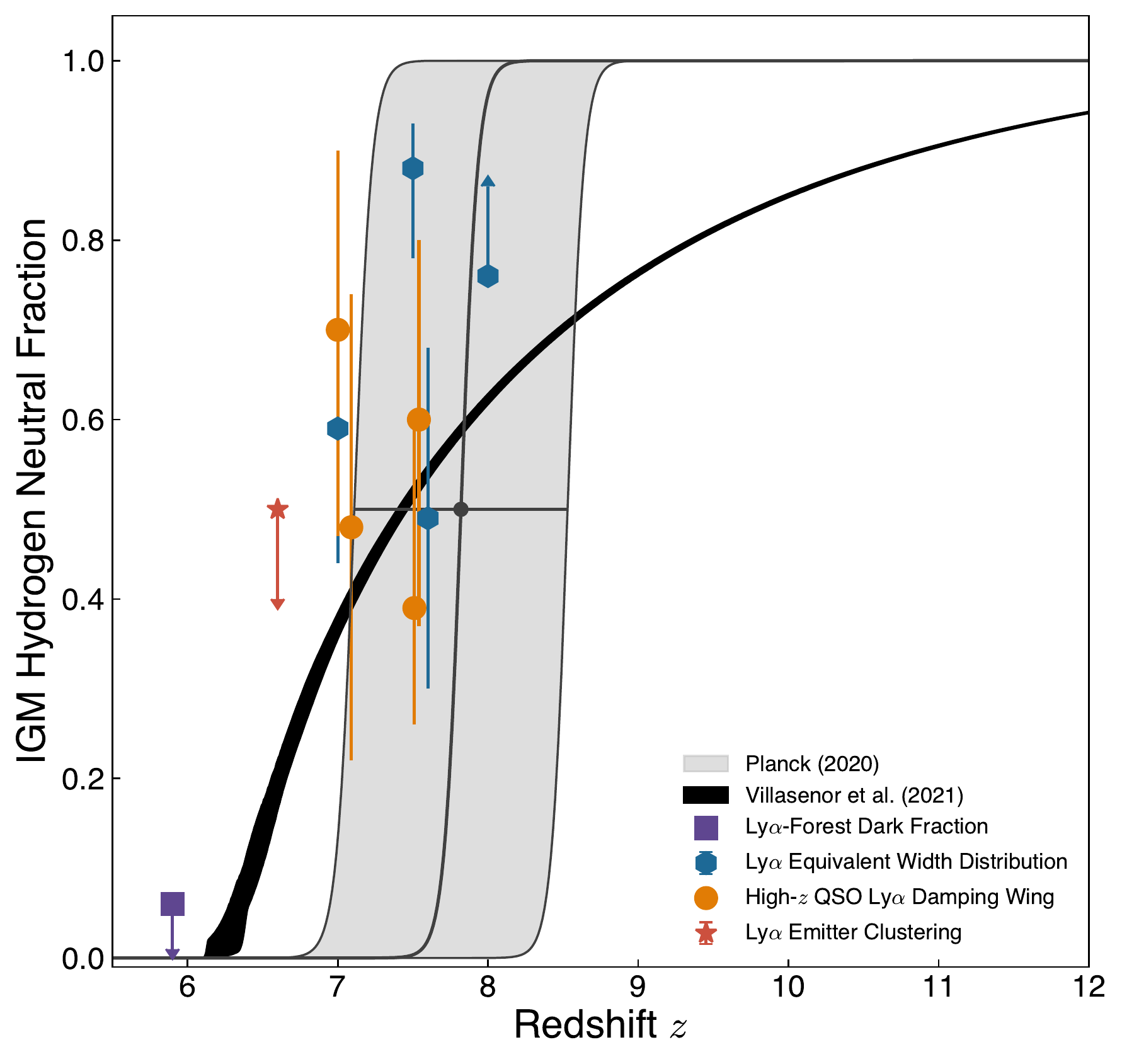}
\caption{Constraints on the
evolution of the hydrogen neutral fraction $\xHI$ in the
intergalactic medium.
Most available constraints result from analyses of
Lyman-$\alpha$ emission and absorption during \reion{}.
From low redshift to high,
constraints are inferred from the fraction of the
Lyman-$\alpha$ and Lyman-$\beta$ forests that are dark
\citep[purple][]{mcgreer2015a}, the clustering of
Lyman-$\alpha$ emitters \citep[red]{ouchi2010a,sobacchi2015a},
the observed equivalent width distribution of $z\sim7-8$
Lyman-$\alpha$ emitters \citep{mason2018a,mason2019a,hoag2019a,jung2020a},
and the Lyman-$\alpha$ damping wing of $z\sim7-8$ quasars
\citep{mortlock2011a,banados2018a,davies2018a,wang2020a,yang2020a}.
Leveraging the sensitivity of the CMB to the electron scattering
optical depth, \citet{planck2020a} infers a
midpoint of reionization of $z(\xHI=0.5)\sim 7.82\pm0.71$ (gray area, reflects the \citealt{planck2020a}
$\tanh$ model for neutral fraction evolution).
Also shown is a prediction from cosmological simulations
for $\xHI$ constrained by the evolution
of the Lyman-$\alpha$ forest transmitted flux power spectrum at $z<6$,
allowing for variations in the range of photo-ionization and
photo-heating history \citep{villasenor2021a}.}
\label{fig:xHI}
\end{figure}

We plot several types of observational
constraints on $\xHI(z)$ and the model by \citet{villasenor2021a}
in Figure \ref{fig:xHI}. 
\citet{mcgreer2015a} quantified the neutral fraction near the
end of reionization using dark regions of the \Lya{} and
\Lyb{} forests in background quasar spectra
that transmit zero flux. At $z=5.9$, this
measurement provides $\xHI\leq0.06\pm0.05$ as is shown
as an upper limit in Figure \ref{fig:xHI}.
The clustering of \Lya{}-emitters limits $\xHI<0.5$ at $z=6.6$
\citep{ouchi2010a,sobacchi2015a}
The equivalent width distribution of \Lya{} emitters at 
$z\sim7-8$ also 
constrains the neutral fraction of the IGM to be 
$\xHI=[0.59_{-0.15}^{0.11}, 0.88_{-0.10}^{+0.05}, 0.49\pm0.19, >0.88]$ at $z=[7,7.5,7.6,8]$
\citep[][]{mason2018a,mason2019a,hoag2019a,jung2020a}.
The \Lya{} damping wing in the spectra of high-redshift quasars
indicates the neutrality of the surrounding IGM, as shown
by \citet{davies2018a}. Four high-$z$ quasars have damping
wing constraints that give 
$\xHI=[0.70_{-0.23}^{0.20}, 0.48_{-0.26}^{+0.26}, 0.60_{-0.23}^{0.20}, 0.39_{-0.13}^{0.22}]$ at $z=[7.00,7.09,7.54,7.51]$
\citep[][]{mortlock2011a,banados2018a,wang2020a,yang2020a}.
Also shown is the constraint from \citet{planck2020a}
on the midpoint of reionization $z(\xHI=0.5)\sim 7.82\pm0.71$,
which assumes a $\tanh$ model for $\xHI(z)$ and is
therefore not independent of the assumed reionization history.
The \citet{villasenor2021a} model is shown for reference, noting
that the model is not fit to the data shown in Figure \ref{fig:xHI}.
Overall, there is general consistency between the data and the
cosmological model, but the relative agreement highlights that
more and significantly tighter constraints would be helpful in
discriminating between different models of the reionization history.

The CMB observations by \citet{planck2020a} provide
an additional constraint on the reionization history.
The optical depth $\tau$ of electron scattering
measured by CMB experiments involves an integral over $\Q(z)$ as
\begin{equation}
\label{eqn:tau}
\tau(z) = c \langle n_H \rangle \sigma_T \int_{0}^{z} f_e \Q(z') H^{-1}(z') (1+z')^2 dz',
\end{equation}
where $c$ is the speed of light, $n_H$ is the comoving hydrogen density, $\sigma_T$
is the Thomson cross-section, $H(z)$ is the Hubble parameter as a function
of redshift $z$, and $f_e$ is the number of free electrons per hydrogen nucleus,
which also depends on the ionization state of helium. 
The {\it Planck} satellite measured a value $\tau=0.0561\pm0.0071$ with
68\% confidence when
constrained by the CMB and low-redshift baryon acoustic oscillations \citep{planck2020a}.
At redshifts where intergalactic hydrogen is fully ionized (a constant $\Q=1$), Equation
1 increases as 
owing to the matter-dominated behavior
$H(z)$ 
and the $(1+z)^2$ weighting of the integrand. Once $\Q$
drops to $\Q<1$, the increase of $\tau$ with redshift moderates.
For a redshift-independent
$\fesc$ and $\xiion$, there is a correspondence between
the Thomson optical depth $\tau$ and the star formation rate density
$\rhoSFR$ \citep{robertson2015a}. For models that match the neutral
fraction evolution shown in Figure \ref{fig:xHI}, the constraint from
$\tau$ mostly then relates to the persistence of the
global cosmic star formation history at very high redshifts ($z>8$).

%
%
\section{MODELS OF HIGH-REDSHIFT GALAXY FORMATION AND REIONIZATION}
\label{sec:models}

Simulations of galaxy 
formation during \reion{} and
its possible connection to the
evolution of the intergalactic medium
have advanced substantially in 
recent years. The inclusion of
radiative transfer (including post-processing)
or radiation hydrodynamics allow for an
understanding of how galaxies produce hydrogen-ionizing
photons, how
those photons escape to the intergalactic
medium, and the bulk effects of the
star-forming population on the
reionization process. Below, we
review some recent efforts to
model the galaxy population before and during
reionization, including simulations that
make specific predictions for observations
with \JWST{}.
Zoom-in cosmological simulations with radiative transfer
provide means for understanding the physical origin of the
escape fraction 
\citep[e.g.,][]{ma2016a,ma2020a,barrow2020a,secunda2020a,mauerhofer2021a},
and these are discussed further in Section \ref{sec:lyc}.

Simulations
with radiative transfer provide a detailed physical
description of \reion{} that the
framework presented in Section \ref{sec:properties}
can only coarsely approximate. These
simulations can track the reionization process
from early times through its completion
and quantify the relative contribution of
faint and bright galaxies to the ionization
budget \citep[e.g.,][]{gnedin2014a,gnedin2014b,xu2016a,lewis2020a,lovell2021a},
Simulations also predict how the
distributions of galaxy metallicities and ages 
influence the mass-dependent contribution of
ionizing photons to the reionization process \citep[e.g.,][]{katz2018a}.
The detailed calculations can follow the 
histories of individual galaxies, and study how the
reionization process affects regions around galaxies
as a function of halo mass \citep{zhu2019a}.

Several recent studies have made predictions
for the photometric properties of high-redshift
galaxies from cosmological simulations. In general,
the simulations roughly match the observed
UV luminosity functions \citep{oshea2015a,gnedin2016a,wilkins2017a,vogelsberger2020a,vijayan2021a}.
Interestingly, despite the observed blue UV continua
of high-redshift galaxies some models predict
substantial obscured star formation even at high-redshift
\citep{vijayan2021a}. Simulations can be used to 
predict the \Lyc{} leakage and nebular emission in
high-redshift galaxies, enabling the interpretation
of future \JWST{} spectra \citep[e.g.,][]{zackrisson2017a}.
Simulations are complemented by
semi-analytical models, which have predicted the
feedback of the reionization process on galaxy
formation \citep{bose2018a}, studied the role of 
AGN in reionization \citep{khaire2016a,dayal2020a}, and
quantified the effects of cosmic variance in 
studies of distant galaxies with \JWST{} \citep{ucci2021a}.

Forthcoming observations with \JWST{} will provide critical
tests of these models. Detailed calculations compute
the redshift-dependent star formation rate and
stellar mass content of galaxies, and until \JWST{} only the
very brightest galaxies in the \reion{} era have had 
stellar mass inferences enabled by \Spitzer{} to compare
with model predictions (see Section \ref{sec:mstar}). By
extending the stellar mass constraints to much
lower galaxy masses, \JWST{}
will determine whether both the instantaneous rate and
time-integrated history of star formation in the
earliest galaxies are reliably predicted by existing theoretical
approaches. Correspondingly,
the model predictions for how stellar mass and star formation
relate to dark matter halo properties can be tested by
\JWST{}. The comparisons between theory and observation will
enable an assessment of our understanding of the physics 
of dark matter structure formation, cooling, and feedback
processes in the first billion years of cosmic history.

%
%
\section{ROLE OF AGN AND QUASARS}
\label{sec:agn}

Active galactic nuclei (AGN) represent a potentially important source 
of ionizing radiation. Their relative rarity and the difficulty
in definitively distinguishing them from galaxies using 
static, broadband photometry make quantifying the role of AGN in the
\reion{} process challenging.
An assessment of the contribution of AGN to \reion{}
requires reviewing important results over the last $\sim$5
years, which have in total led to the current near-consensus that
AGN are a subdominant source of hydrogen ionizing photons 
during the reionization era. Nonetheless,
high-redshift quasars can provide
critical probes of the intergalactic medium neutrality during
\reion{}.

\citet{giallongo2015a} identified faint AGN candidates in
\HST{} CANDELS GOODS-S imaging, leading to estimates of
enhanced numbers of low-luminosity AGN to the hydrogen
ionizing budget at high-redshift. \citet{madau2015a}
carefully considered the implications of these potential
AGN on reionization and concluded that if a large number
of faint AGN existed at $z\sim6$ they could potentially
provide enough Lyman continuum radiation to reionize
the universe without a substantial contribution from 
galaxies. The escape fraction of ionizing photons
from AGN should be substantially enhanced relative
to galaxies, with estimates at $z\sim3-4$ of $\approx75\%$
\citep{grazian2018a} and, even with their relative rarity
compared with galaxies, faint AGN would be efficient
agents of reionization \citep[e.g.,][]{khaire2016a}.

Later studies of potential AGN contributions to reionization
drew more conservative conclusions about their role.
If AGN did contribute most of the high-redshift Lyman continuum,
it would lead to large variations in the hydrogen \Lya{} forest
during the end of \reion{}, impact the mean opacity of the HeII
\Lya{} forest, and lead to substantial photo-heating of the IGM
through early HeII reionization \citep{daloisio2017a}. 
\citet{parsa2018a} reported that many potential faint
AGN in \HST{} images were unconvincing and, given
candidates they found reliable, that AGN failed by
a factor of roughly an order of magnitude at redshift $z\sim6$
to solely reionize intergalactic hydrogen. Subaru
searches for bright quasars at $5.7\lesssim z \lesssim 6.5$
\citep{matsuoka2016a} produced AGN luminosity functions that
likely would only support 10\% of the required ionizing 
photon budget at the end of reionization \citep{matsuoka2018a}.
Recent determinations of the high-redshift AGN luminosity
function agree that at $z\sim6$ AGN are subdominant \citep[$\sim$ a few percent;][]{kulkarni2019a},
and current uncertainties in the faint end of the high-redshift bolometric luminosity
are unlikely to change this conclusion \citep{shen2020a}.
Given the current observed AGN luminosity functions, most theoretical
models that can reproduce them conclude that AGN do not drive
reionization \citep[e.g.,][]{hassan2018a,mitra2018a} but that
AGN can dominate the Lyman continuum production among the galaxy
population at the highest stellar masses \citep{dayal2020a}.

Some of our most critical information on the ionization state of the
intergalactic medium is inferred from the spectra of highest-redshift
quasars. \citet{mortlock2011a} discovered the first quasar at 
$z>7$ in the UKIDSS Large Area Survey (ULAS), with redshift $z=7.085$.
This discovery has been followed by several others at $z\approx7.0-7.642$
identified by a combination of ULAS, Dark Energy Survey, WISE, UKIRT, and
Subaru observations \citep{banados2018a,yang2019a,yang2020a,wang2020a,izumi2021a,wang2021a},
as well as interesting QSOs at just slightly lower redshifts \citep[e.g.,][]{wangf2019a}.
\citet{davies2018a} used the \Lya{} damping wing profile in the spectra of the
quasars discovered by
\citet{mortlock2011a} and \citet{banados2018a} to place constraints on the
neutral fraction of the IGM at $z=7.085$ ($\xHI=0.48_{-0.26}^{+0.26}$) 
and $z=7.54$ ($\xHI = 0.60_{-0.23}^{+0.20}$, respectively. Two
additional quasars have had \Lya{} damping wing measurements of 
their local IGM neutral fraction at $z=7.00$ \citep[$\xHI=0.70_{-0.23}^{+0.20}$;][]{wang2020a}
and $z=7.515$
\citep[$\xHI=0.39_{-0.13}^{+0.22}$;][]{yang2020a}.
These four measurements are discussed in the context of the
evolving neutral fraction in Section \ref{sec:Q}.
Unfortunately, the high-redshift quasar at $z\sim7.642$ discovered by \citet{wang2021a}
cannot currently provide a reliable \Lya{} damping wing measurement owing to
complications from Si IV broad absorption lines.

Combined observations of AGN and galaxies can also provide insight into
\reion{}.
Quasars create ionized proximity zones, and the study of galaxies in the
proximity zones can inform us about the end of the reionization process 
\citep{bosman2020a}. Along the line of sight to high-redshift quasars,
the correlation between IGM transmission and the presence of galaxies
near the sightline can enable estimates of the high-redshift galaxies' Lyman
continuum escape fractions \citep{kakiichi2018a}, which provide
$\fesc = 0.14_{-0.05}^{+0.28}$ \citep[][see also Section \ref{sec:lyc}]{meyer2020a}.

%
%

\section{INNOVATIONS IN HIGH-REDSHIFT GALAXY OBSERVATIONS}
\label{sec:innovations}

The pursuit of high-redshift galaxies during \reion{} has involved primarily the combination of deep multiband imaging with drop-out selections and follow-up spectroscopy in contiguous fields chosen for their accessibility, reddening, and some sense of how typical they might be for regions of the distant universe. Over the last decade, significant innovations in the search for high-redshift galaxies have been made and some are described below. These
innovations should not be seen as somehow separate from the standard approach, and indeed the results from these observations are integrated in the discussion elsewhere in this review (especially in Section \ref{sec:rhoUV}). Nonetheless, they warrant particular note as each innovation has relevance for future studies of the high-redshift universe with \JWST{}.

\subsection{Gravitationally-Lensed Observations}
\label{sec:lensing}

The use of gravitational lenses to magnify faint objects has become increasingly widespread
in galaxy surveys.
Significant investment by \HST{} through the Frontier Fields program \citep{lotz2017a}
has now provided deep multiband images of several strong-lensing clusters of galaxies (Abell 2744, MACSJ0416.1-2403, MACSJ0717.5+3745, MACSJ1149.5+2223, Abell S1063, and Abell 370). 
The Frontier Fields complemented the
predecessor Cluster Lensing and Supernova Survey with Hubble (CLASH) imaging survey \citep{postman2012a}
and the Grism Lens-Amplified Survey from Space \citep[GLASS;][]{treu2015a} that provided \HST{} IR grism
observations of the Frontier Fields and four additional clusters.
While the Frontier Fields focused on deep imaging of select clusters and parallel ``blank''
fields, the Reionization Lensing Cluster Survey \citep[RELICS;][]{coe2019a} targeted
a larger number (41) of clusters with shallower imaging.
The cluster observations by the Frontier Fields, CLASH, GLASS, and RELICS
can probe to substantially fainter luminosities at fixed integration time than 
blank field observations that do not benefit from the aid of gravitational
lensing. However, this increased power is balanced against the added
complexity of modeling the gravitational lens systems to understand
the magnification map of the foreground \citep[e.g.,][]{meneghetti2017a}
and by the decreased volume
probed at high-magnification relative to a blank field at the same
intrinsic depth that leads to higher cosmic variance uncertainty \citep{robertson2014a}.

Gravitational lensing clusters have been particularly useful for
identifying distant dropout galaxies \citep[e.g.,][]{zheng2012a,zheng2017a,salmon2018a,salmon2020a}
investigating the faint-end slope of the galaxy luminosity
function during \reion{}
\citep{atek2015a,mcleod2015a,mcleod2016a,livermore2017a,bouwens2017a,ishigaki2018a,yue2018a,kawamata2018a,bhatawdekar2019a,furtak2021a}.
Lensing clusters have also been helpful in amplifying signals for
reionization-era observations that are less efficient than \HST{} broadband imaging,
including \Spitzer{} IRAC \citep{shipley2018a,strait2020a,kikuchihara2020a}, 
slit and grism spectroscopy
\citep[e.g.][]{schmidt2016a,mason2019a,pelliccia2020a,fuller2020a}, and 
intermediate band imaging \citep{hernan-caballero2017a}.
The results of many of these efforts have been incorporated into
the analyses that arrive at the luminosity functions reported
in Section \ref{sec:rhoUV} \citep[e.g.,][]{bouwens2021a}.

\subsection{Pure-Parallel Surveys}

The ability with \HST{} and with the forthcoming \JWST{} to integrate
with two instruments simultaneously provides an extreme advance
in observational efficiency. Pure-parallel surveys enable extremely
wide areas to be covered essentially for free while prime focus
surveys are being conducted, allowing for survey volumes that
exceed traditional contiguous campaigns at the expense of some
added complexity of analysis and non-uniformity in the depths
and coverage. Two important \HST{} parallel imaging surveys
include the Brightest of Reionizing Galaxies (BoRG) Survey \citep{trenti2011a}
and the Hubble Infrared Pure Parallel Imaging Extragalactic Survey \citep[HIPPIES;][]{yan2011a}.

Pure-parallel imaging provides some of the best constraints
on rare, luminous high-redshift galaxies \citep{bradley2012a,schmidt2014a,calvi2016a,morishita2018a,bridge2019a,rojas-ruiz2020a,morishita2020a,morishita2021a}.
In discovering luminous objects, these surveys also provide
candidate high-redshift galaxies detected in parallel
grism observations \citep{bagley2017a} or that can
be targeted
spectroscopically to help determine the reliability of 
drop-out selections \citep{livermore2018a}.

\subsection{\ALMA{} Observations of High-Redshift Galaxies}

A remarkable advance in the study of \reion{} has been
the ability to detect and interpret properties of gas and
dust in
distant galaxies. With exquisite sensitivity, \ALMA{} 
can reveal continuum and spectroscopic signatures deep into the
reionization era that have been unobtainable in the past.
These observations complement the photometry and spectroscopy
from \HST{}, ground-based large telescopes, and \JWST{} that
remain limited to much shorter wavelengths than what
\ALMA{} can achieve.

In addition to the critical \ALMA{} imaging of the 
Hubble Ultra Deep Field \citep{dunlop2017a}, spectroscopic
surveys of the HUDF and other deep fields have been conducted with \ALMA{}
\citep{hodge2013a,walter2016a,franco2018a}. \ALMA{} spectroscopy
allows for the detection of highly
redshifted [OIII] $88\um$ \citep[e.g.,][]{inoue2016a} and [CII] $158\um$ \citep[e.g.,][]{maiolino2015a}
lines during \reion{}, and
directly probes coolants in the interstellar medium of
distant galaxies.
The spectroscopic confirmation of the highest-redshift galaxies with \ALMA{}, especially
with [OIII] \citep[e.g.][]{hashimoto2018a}, is
discussed in Section \ref{sec:specz}. The longer-wavelength [CII] line is now often used
to study $z\sim7$ galaxies \citep{pentericci2016a,bradac2017a,carniani2017a}.
Ongoing large programs with ALMA will study many bright blank field galaxies
in both [CII] and [OIII] \citep{bouwens2021b}.
The high-resolution of \ALMA{} allows for the investigation of 
spatial offsets between \Lya{} and [CII] emission \citep{maiolino2015a} at redshifts $z>7$,
providing new information on the interior structure of high-redshift galaxies.
\ALMA{} continuum measurements also provide evidence of dust in at least some $z>7$ objects
in abundances not uncommon for present-day galaxies \citep[][see \citealt{casey2014a} for a broader review of dust in high-redshift galaxies]{watson2015a,laporte2017a}. 
Comparisons between observational and theoretical
studies of the connection between [CII] emission and star formation during
\reion{} will provide insight into the evolving stellar radiation fields and conditions
inside these early-forming galaxies \citep{lagache2018a}.

\section{HIGH-REDSHIFT GALAXY STELLAR MASS CONSTRAINTS}
\label{sec:mstar}

The stellar mass content of high-redshift galaxies remains an
area with great promise for future study, especially
with \JWST{}. To date, the best constraints on the
stellar mass of high-redshift galaxies come from 
combinations of \Spitzer{} IRAC $3.6\um$ and
$4.5\um$ observations with \HST{} or ALMA data. For a summary
of \Spitzer{} results at $z>4$, see the recent review
by \citet{bradac2020a}.

Once the HUDF, CANDELS, and CLASH
survey data had been completed and augmented by \Spitzer{}
data \citep[e.g.,][]{bradac2014a,labbe2015a}, combining
\Spitzer{} with \HST{} WFC3/IR enabled the 
first constraints on the stellar mass function evolution
to redshift $z\sim7$ \citep{stark2013a,duncan2014a,grazian2015a}.
These studies found that the high-redshift stellar mass function
steepened considerably to $z\sim7$ \citep[see also][]{davidzon2017a}.
By redshift $z\sim8$ the stellar mass function reaches $\alpha<-2$, but
with large uncertainties \citep{song2016b}. Similar conclusions
were drawn by converting the rest-frame optical luminosity
function to $z\sim7$ to a stellar mass function \citep{stefanon2017a}
and by using initial observations of the Frontier Fields to extend this
observed steepening to $z\sim9$ \citep{bhatawdekar2019a}. 
Subsequently, \citet{kikuchihara2020a} have claimed based on more
complete Frontier Fields data that $\alpha\approx-1.8$, perhaps 
moderating the abundance of low stellar mass objects during \reion{}.
Recently, with expanded \Spitzer{} imaging of the GOODS fields,
\citet{stefanon2021a} have been able to constrain the stellar
mass of objects to $10^{8}\Msun$ with $\approx70\%$ completeness.
They find stellar mass functions to $z\sim10$ consistent with
a steep faint-end slope $\alpha\approx-2$, and that while the
stellar mass in galaxies increases by $\sim1000\times$ over
the redshift range $z\sim6-10$ the stellar mass to halo mass
($M_{\star}-M_{h}$) ratios of galaxies do not substantially change. Stellar mass
and halo mass appear to be growing jointly apace during \reion{},
instilling the $M_{\star}-M_{h}$ relation observed at lower
redshifts \citep[for a review, see][]{wechsler2018a}.

Despite this progress, the role for \JWST{} will be substantial.
A key theme in the interpretation of stellar mass observations is
the possible role of nebular emission contamination of the rest-frame
optical data probed by \Spitzer{} \citep[e.g.,][]{stark2013a,smit2014a}.
Owing to the strong nebular lines powered by reprocessing of
internal Lyman continuum radiation, galaxies can display boosted
broadband measures of their rest-frame optical emission beyond
the continuum level supplied by evolved stars. Depending on the
redshift and line strength, as strong lines pass through the \Spitzer{} bands
the rest-frame optical broadband colors of the object may change
\citep[e.g.,][]{roberts-borsani2016a}.
Without spectroscopic redshifts, a wide range of stellar masses
can be inferred from
\Spitzer{} IRAC observations of even
bright $z\sim8$
candidate galaxies \citep{strait2020a}. \JWST{} will provide
rest-frame optical spectroscopy to $\lambda\approx5\um$ with
NIRSpec at depths sensitive enough to reach [OIII]+H$\beta$
in $z\sim8-9$ galaxies, and can disentangle the possible
contribution of nebular emission and evolved stellar populations
to the rest-frame optical broadband flux observed.

While the ultimate resolution to this issue may be 
rest-frame optical spectroscopic observations
 with \JWST{} to $z\sim9$,
in the near term spectroscopic redshift confirmation with
ALMA provides a significant benefit in the stellar mass
solution of high-redshift objects \citep{roberts-borsani2020a}.
Observations of 
the gravitationally-lensed galaxy MACS1149-JD, originally
identified in CLASH by \citet{zheng2012a}, with ALMA
revealed $88\um$ [OIII] at $z=9.1096$ \citep{hashimoto2018a}.
This redshift confirmation enabled the stellar population
synthesis models to infer confidently the presence of
evolved stellar mass in this distant object, pointing to
star formation commencing as early as redshift $z\sim15$.
Without the spectroscopic redshift, the possibility of
nebular contamination would make the stellar mass inference
less certain.
Ground-based \Lya{} redshifts can still provide a similar
benefit in firming up the stellar population inferences on
$z\sim9$ galaxies, which has allowed
\citet{laporte2021a} to determine that in $z\gtrsim9$
galaxies about $70\%$ of their stellar mass may 
be in place before $z\sim10$.
Regardless of the current facility enabling the
observation, 
depending on the exact
redshift, in some cases
nebular contamination may not be possible to
rule out given pre-\JWST{} data.

%
%

\section{SPECTROSCOPY OF THE HIGHEST-REDSHIFT GALAXIES}
\label{sec:spec}

Spectroscopy enables us to understand the
astrophysical principles that give rise to the observable
properties of the galaxy population. While we
can learn much from the photometric properties
of galaxies alone, the detailed wavelength-dependent
flux information that is available from
spectra allows us to more closely 
connect observations with atomic and nebular
physics. 
For the distant universe, \JWST{} will
measure rest-frame optical spectra that
contain rich physical information on the
detailed properties of galaxies. \JWST{} will
also enable the routine spectroscopic confirmation
of photometric candidate sources at extreme
distances. Below, we review some of the current
status of redshift confirmation and spectroscopic
astrophysics for distant galaxies. Anticipated
advances with \JWST{} in these areas will be discussed in
Section \ref{sec:jwst}.

\subsection{Spectroscopic Confirmation of Distant Galaxy Redshifts}
\label{sec:specz}

The large majority of the galaxies contributing to determinations of
the evolving luminosity density are photometric detections only, with
substantial uncertainty in their redshifts.
The capabilities of facilities like Keck Observatory, the Very Large
Telescope, and Magellan to provide deep, near-infrared spectroscopy from the ground
and \HST{} to enable slitless infrared spectroscopy without contamination
by the sky
have substantially advanced our understanding of the high-redshift universe
by observing Ly$\alpha$ ($\lambda=1216$\AA),
NV ($\lambda=1243$\AA) 
CIV ($\lambda=1548,1550$\AA),
OIII] ($\lambda=1666$\AA),
and CIII] ($\lambda=1907,1909$\AA) emission lines at $z>7$.
Now, combined with the powerful ability of the Atacama Large Millimeter/submillimeter
Array (ALMA) to detect redshifted [OIII] ($\lambda=88\um$) and [CII] ($\lambda=157\um$),
spectroscopic observations of galaxies at $z>7$ have become increasingly
reliable and informative.

Redshifted Ly$\alpha$ remains a mainstay for high-redshift spectroscopic
confirmations. 
A small population of galaxies with redshifts $7.5\gtrsim z\gtrsim8$ 
have been confirmed, often selected from \HST{} CANDELS imaging.
These include an early discovery 
of \Lya{} emission at $z=7.51$ \cite{finkelstein2013a}, which
also has a grism continuum measurement of the Lyman break
\citep{tilvi2016a}. Over the next few years, this
object was followed by a range
of \Lya{} emitters spectroscopically confirmed at $z=7.45-8$
\citep{oesch2015a,roberts-borsani2016a,song2016a,stark2017a,hoag2017a,larson2018a,jung2019a},
including a gamma-ray burst host galaxy at $z=7.8$ \citep{tanvir2018a} and a primitive galaxy group at $z=7.7$ \citep{tilvi2020a}.
Recently, a large-scale campaign with Keck MOSFIRE using 10 nights
targeted six dozen systems, achieving 10 detections at redshifts
$z=7.1-8.2$ \citep[][see also \citealt{pentericci2018a}]{jung2020a}.

The demonstrated ability to spectroscopically confirm
high-redshift galaxies using near UV emission lines
redward of \Lya{} was a hallmark achievement for the
study of early galaxies. As discussed in Section \ref{sec:Q},
the presence of neutral hydrogen in the IGM during the
\reion{} process can attenuate substantially \Lya.
Having other available lines for spectroscopic redshift
confirmation may therefore be critical during the
reionization proceess and at earlier times.
\citet{stark2015b} detected CIII] $1909$\AA{} in galaxies
at $z\sim6.029$ and $z=7.213$ with known \Lya{} 
redshifts. \citet{stark2015a} detected CIV $1243$\AA
in a galaxy with known \Lya{} redshift $z=7.045$.
\citet{laporte2017a} detected \Lya{} and CIII] in
a $z\sim6.8$ galaxy. 
At slightly lower redshift, an interesting
discovery for understanding the internal ionizing
spectrum of star-forming galaxies was the
confirmation of CIV $1550$\AA{} and 
OIII] $\lambda=1666$\AA{} in a galaxy with
known \Lya{} $z=6.11$ \citep{mainali2017a}.
The presence of CIII] has now been 
confirmed in galaxies at $z=7.51$ 
\citep[][same object as \citealt{finkelstein2013a}]{hutchison2019a},
$z=7.73$ \citep[][same object as \citealt{oesch2015a}]{stark2017a},
and recently $z=7.945$ \citep{topping2021a}.

Confirming spectroscopically the redshifts of the
most distant candidates proves especially useful. Given the
rapid apparent decline in the UV luminosity density
$\rhoUV$ in photometrically-selected candidates $z>8$, any confirmed 
redshifts further establish what galaxy population is
reliably known to exist at these early times.
Extremely distant objects have been confirmed
via \Lya{}  and NV $1243$\AA{} emission
at redshift $z=8.68$ \citep{zitrin2015a,mainali2018a},
and with
\Lya{} and ALMA [OIII] $88\um$ at $z\sim8.38$ \citep{laporte2017a}.

The object MACS0416-Y1 identified in the \HST{} Frontier
Fields \citep{zheng2012a} was spectroscopically confirmed 
at $z=8.312$ with ALMA
detections of [OIII], [CII],
and dust emission \citep{tamura2019a,bakx2020a}.
Using Keck and VLT, \citet{laporte2021b} confirmed
\Lya{} redshifts of $z=8.78$ for GN-z-10-3 and
(tentatively) $z=9.28$ for  MACS0416-JD.
The candidate MACS1149-JD1 was confirmed to lie at
$z=9.1096$ using ALMA [OIII] $88\um$  \citep{hashimoto2018a}.
Interestingly, in the $z=8.38$ and $z=9.1$ [OIII] $88\um$ 
emitters, [CII] $158\um$ has not yet been detected \citep{laporte2019a}.
Some objects at lower redshifts ($z=7.15$) have simultaneous 
ALMA [OIII] $88\um$, [CII] $158\um$, and dust continuum detections
\citep{hashimoto2019a}.

The most distant known, spectroscopically-confirmed object is 
the GOODS-N galaxy GN-z11 originally
identified using \HST{} WFC3/IR imaging. 
\citet{oesch2016a} used \HST{} grism spectroscopy to measure a 
continuum spectral break for GN-z11 corresponding
to redshift $z=11.09^{+0.08}_{-0.12}$. Recently,
using \Keck{}/MOSFIRE \cite{jiang2021a} confirmed the
redshift using three restframe-UV emission lines, finding a
redshift of $z=10.957$ if the lines correspond to
the CIII] doublet and OIII].

\subsection{Spectroscopic Probes of Reionization-Era Galaxy Astrophysics}
\label{sec:spec_astro}

Beyond the utility of redshift confirmation
for distant objects, spectra contain vital
information on the astrophysics of early galaxies.
Given the current limitations on obtaining
rest-frame optical spectra of distant galaxies
before \JWST{}, most spectroscopy at $z\gtrsim3$
is limited to rest-frame UV. The \Lya{} emission
line itself provides rich astrophysical information,
but we defer to the recent review by \citet{ouchi2020a} on
\Lya{} emitters and their properties. Below, we touch
briefly on the use of other lines as potential
astrophysical probes of high-redshift galaxies.

Given their current accessibility via ground-based
spectroscopy, UV emission lines are of particular
interest today. 
\citet{jaskot2016a} used photoionization models to 
show that CIII] is expected to be the second
brightest emission line (next to \Lya) in high-redshift
galaxies blueward of $\lambda\sim2700$\AA{} in the rest-frame.
\citet{stark2015b} and \citet{stark2017a} have connected
the detection of CIII] in $z>7$ galaxies to enhanced internal
ionization fields. Not all high-redshift \Lya{} display
bright UV metal lines, or show evidence for NV $1243$\AA{}
that may indicate an AGN contribution or strongly shocked ISM 
gas \citep{mainali2018a}.

Once \JWST{} NIRSpec provides the ability to expose
deeper and redder spectra than currently possible, a host of additional
probes of high-redshift galaxies will become available.
At redshifts $z\sim5$ and $z\sim6$, the mean
rest-UV spectra of star-forming galaxies show evidence
for enhanced photon production rates relative to
local galaxies \citep{pahl2020a,harikane2020a}.
The  presence of strong [OIII]+H$\beta$
emission in high-redshift galaxies inferred
from \Spitzer{} photometry and colors will provide
additional information. High-redshift objects with strong
inferred [OIII] equivalent widths show abundant
\Lya{} emission, suggesting a strong 
radiation field \citep{endsley2021a,endsley2021b}.
The [OIII] emission itself requires efficient ionizing photon
production, and photometric measures of OIII+Hbeta suggest
this will be prevalent at $z\sim5-8$ \citep{castellano2017a,harikane2018a,debarros2019a}. \JWST{} will address all these issues
with deeper, redder spectra.

%
%
\section{FUTURE PROGRESS WITH \JWST{}}
\label{sec:jwst}

Given its unprecedented ability to study the
distant universe,
\JWST{} will undoubtedly revise our understanding of 
\reion{}. To explain how \JWST{} will revolutionize
studies of early galaxy formation, we review both
the powerful \JWST{} instruments and a subset of the incredible array
of Director's Discretionary-Early Release Science (DD-ERS),
Guaranteed Time Observations (GTO), and Cycle 1 Guest Observers
(GO) programs planned for the first year of \JWST{} operations.

\subsection{Overview of \JWST{} Instrumentation}

\JWST{} provides four ground-breaking instruments, each
with application to \reion{} science. We briefly
describe the array of instrumentation below, providing
information on the field of view, pixel scale,
wavelength coverage, observing modes, and typical 
sensitivities. For reference,
$M^{\star}\approx-21$ galaxies at $z\sim8-9$ have
$f_{\nu}\approx100$ nJy at $\lambda_{obs}\approx1.5\um$,
stellar masses of $M_{\star}\approx10^{9}\Msun$
\citep[e.g.,][]{stefanon2021a},
and UV line fluxes of
$f\approx1-5\times10^{-18}\mathrm{erg}\,\mathrm{s}^{-1}\,\mathrm{cm}^{-2}$ \citep[e.g.,][]{laporte2021b,topping2021a}.

\subsubsection{MIRI}
\label{sec:miri}

The Mid-Infrared Instrument \citep[MIRI;][]{rieke2015a,wright2015a}
extends the capabilities of \JWST{} to wavelengths
$\lambda\approx4.9-28.8\um$. MIRI includes imaging \citep{bouchet2015a}, coronagraphic
\citep{boccaletti2015a},
and spectroscopic \citep{wells2015a,kendrew2015a} observing modes.
The MIRI detectors are arsenic-doped
silicon impurity band conduction devices \citep[][]{rieke2015b}.
The imaging mode has a 0.11''/pixel plate scale over a 74'' $\times$ 113'' field of view.
There are eight broad imaging filters including
\jf{F560W} ($\lambda\approx5.6\um$),
\jf{F770W} ($\lambda\approx7.7\um$),
\jf{F1000W} ($\lambda\approx10.0\um$),
and long-wavelength channel wide filters
\jf{F1280W} ($\lambda\approx12.8\um$),
\jf{F1500W} ($\lambda\approx15.0\um$),
\jf{F1800W} ($\lambda\approx18.0\um$),
\jf{F2100W} ($\lambda\approx21.0\um$), and
\jf{F2550W} ($\lambda\approx25.5\um$).
\citet{glasse2015a} presented the predicted
sensitivity of MIRI in its various observing
modes. For reference, $\lambda=7.7\um$ (\jf{F770W}) imaging
reaches $SNR\approx10$ point source sensitivities of
$f_\nu\sim 250$nJy in $t=10$ks (see their Table 3).

\subsubsection{NIRCam}
\label{sec:nircam}

The Near Infrared Camera \citep[NIRCam;][]{rieke2005a}
features two HgCdTe detector modules with dichroics to expose
short ($\lambda0.6-2.3\um$; 0.031''/pix) and long
($\lambda=2.4-5.0\um$; 0.063''/pix) wavelength channels 
in parallel. The two detector modules
cover 2.2' $\times$ 2.2' areas (9.7 arcmin$^{2}$ total).
NIRCam also enables wide-field slitless spectroscopy
at $\lambda=2.4-5.0\um$ with $R\sim1600$, coronagraphy,
and photometric and grism time-series observations.

There are 29 NIRCam filters available. For high-redshift
photometric studies of faint sources, the short-wavelength
channel wide filters
\jf{F090W} ($\lambda\approx0.90\um$),
\jf{F115W} ($\lambda\approx1.15\um$),
\jf{F150W} ($\lambda\approx1.50\um$),
\jf{F200W} ($\lambda\approx2.00\um$)
and long-wavelength channel wide filters
\jf{F277W} ($\lambda\approx2.77\um$),
\jf{F356W} ($\lambda\approx3.56\um$) and
\jf{F444W} ($\lambda\approx4.44\um$)
will be frequently used. These filters
are complemented by the short-wavelength
medium filters
\jf{F182M} ($\lambda\approx1.82\um$),
\jf{F210M} ($\lambda\approx2.10\um$)
and long-wavelength medium filters
\jf{F335M} ($\lambda\approx3.35\um$), and
\jf{F410M} ($\lambda\approx4.10\um$).
There are a very large number of additional filters
that make NIRCam extremely versatile.
NIRcam provides extremely sensitive photometry,
reaching $SNR\approx10$ for $f_{\nu}=10$nJy point
sources in $t\approx10$ks with \jf{F200W}.

\subsubsection{NIRISS}
\label{sec:niriss}

The Near Infrared Imager and Slitless Spectrograph
\citep[NIRIS;][]{doyon2012a} provides
wide-field slitless spectroscopy (WFSS; $\lambda=0.8-2.2\um$ with $R\sim150$ 
over 133'' $\times$ 133''), single object slitless spectroscopy
($\lambda=0.8-2.2\um$ with $R\sim700$) in six filters,
aperture masking
interferometry, and parallel imaging in
twelve filters. NIRISS features two grisms that disperse
in perpendicular directions, allowing for deblending
of spectra from complex source scenes.
The NIRISS pixel scale is 0.065''/pix.
Beyond $\lambda\approx1.3\um$, the NIRISS WFSS
$SNR\approx10$ line sensitivity
is $F\sim10^{-17}$ ergs s$^{-1}$ cm$^{-2}$ in $t=10$ks.
The point source imaging $SNR\approx10$
sensitivity is $f_\nu\approx9.7$nJy in $t=10$ks for
\jf{F200W}.

\subsubsection{NIRSpec}
\label{sec:nirspec}

The Near Infrared Spectrograph \citep[NIRSpec;][]{ferruit2012a}
provides prism, medium, and high-resolution spectroscopy at
$\lambda\sim1-5\um$. NIRSpec features
a micro-shutter assembly (MSA) consisting of hundreds of
thousands of magnetically actuated slitlets. The MSA
enables hugely multiplexed ($\sim$hundreds of simultaneous targets) 
multi-object spectroscopy over a 3.6' $\times$ 3.4' field of view,
providing a huge gain in efficiency for restframe-UV and optical 
spectroscopy of distant sources.
The instrument also features five fixed slits, a 3'' $\times$ 3'' integral field unit, and a time-series spectrograph.

There are nine disperser-filter combinations. Four
combinations provide $R\sim1000$ spectroscopy over
$\lambda=0.7-1.27\um$ (\jf{G140M}/\jf{F070LP}),
$\lambda=0.97-1.84$ (\jf{G140M}/\jf{F100LP}),
$\lambda=1.66-3.07$ 
(\jf{G235M}/\jf{F170LP}),
and $\lambda=2.87-5.10\um$ (\jf{G395M}/\jf{F290LP})
wavelength ranges. The four
high-resolution disperser-filter combinations
enable 
$\lambda=0.81-1.27\um$ (\jf{G140H}/\jf{F070LP}),
$\lambda=0.97-1.82$ (\jf{G140H}/\jf{F100LP}),
$\lambda=1.66-3.05$ 
(\jf{G235H}/\jf{F170LP}),
and $\lambda=2.87-5.14\um$ (\jf{G395H}/\jf{F290LP})
spectroscopy at $R\sim2700$.
The \jf{PRISM}/\jf{CLEAR} combination provides
$R\sim100$ spectroscopy over the wide wavelength
range $\lambda=0.60-5.30\um$. 

\subsection{JWST Programs}
\label{sec:cycle1}

\begin{figure}[h]
\includegraphics[width=\linewidth]{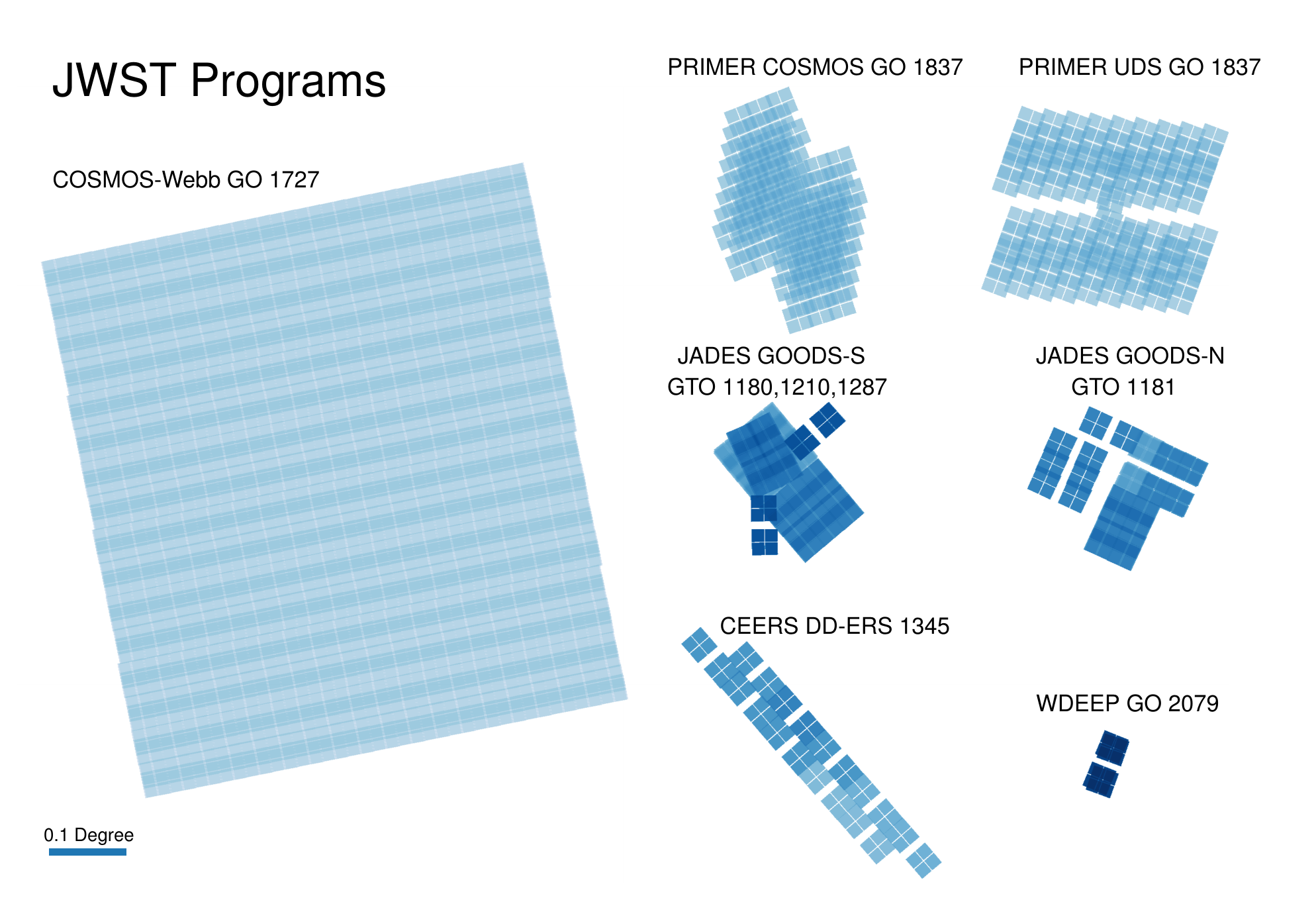}
\caption{Approximate NIRCam prime and parallel
exposure maps for a
subset of extragalactic programs planned during
Cycle 1 with \JWST{} relevant for studies of \reion{}.
These programs include COSMOS-Webb (GO 1727), JADES
GOODS-N (GTO 1181) and GOODS-S (GTO 1180, 1210, 1287 shown; 
GTO 1286 parallels omitted) fields,
PRIMER COSMOS and UDS fields
(GO 1837), CEERS (DD-ERS 1345) in the EGS field, 
WDEEP (GO 2079) on UDF Parallel 2. The maps are extracted
from the public APT data available for each program and
represent the relative exposure time and coverage of 
their {\it F115W} or {\it F200W} imaging.}
\label{fig:jwst}
\end{figure}

During Cycle 1, \JWST{} will
execute a wide range of observing programs
focused on extragalactic science. Here, we
discuss an incomplete subset of the DD-ERS, GTO, and
GO programs relevant
for studies of the distant universe.
Figure \ref{fig:jwst} shows approximate
NIRCam \jf{F115W} or \jf{F200W}
exposure maps for a subset of these DD-ERS,
GTO, and GO programs computed from their public
Astronomer's Proposal Tool (APT) files. Each
program in Figure \ref{fig:jwst} is reviewed
below. In each case, the programs are presented in
order of their Program ID and information about
the observations summarized from the APT
files and Phase2 Public abstracts. The relative
areas of the observed fields should be accurate,
and the color scale indicates the relative 
exposure times in each field.

\subsubsection{Director's Discretionary Early Release Science Programs}
\label{sec:dd-ers}

The substantial investment of
public Director's Discretionary-Early Release Science (DD-ERS) programs will provide an important
community resource of public \JWST{} data that exercise the
instrumentation.
For \reion{} science, the DD-ERS programs that expose
early deep imaging in fields with supporting ancillary
data, that provide demonstrations of the \JWST{} instrument
performance on fields with key extragalactic targets, or
that showcase the new infrared spectroscopy capabilities 
of \JWST{} on objects of interest will prove most useful.
Here, we describe two \JWST{} DD-ERS programs particularly
relevant for \reion{} research.

\jp{Through the Looking GLASS} \citep[DD-ERS 1324; 35.1h;][]{glass1324} will
expose NIRISS WFSS, NIRSpec MSA, and NIRCam
on the Frontier Fields gravitational lensing
cluster Abell 2744. The NIRISS and NIRSpec prime 
observations cover a single
pointing, with the NIRISS data probing $\lambda=1-2.2\um$ with the
$R=150$ grism ($\mab\approx25.5-26.3$ per pixel) and
with direct images ($\mab\sim27.5-27.8$).
Their NIRSpec observations use the high-resolution gratings at $\lambda\approx0.97-5.14\um$
at $\texp\approx17.7$ks per band.
The NIRCam parallel observations cover two nearby fields
in \jf{F090W}, \jf{F115W}, \jf{F150W}, \jf{F200W}, \jf{F277W}, \jf{356W},
and \jf{F444W} with exposure times weighted to reach $\mab \approx29-29.4$
in each filter.

The \jp{Cosmic Evolution Early Release Science (CEERS)} program
\citep[DD-ERS 1345; 65.7h;][]{ceers1345}
will
observe $\approx100$~\sqarcmin{} in the Extended Groth
Strip with NIRCam and MIRI imaging,
and NIRSpec MSA and NIRCam WFSS spectroscopy.
CEERS will conduct parallel observations, exposing in 
NIRCam with parallel imaging while prime observing with 
NIRSpec or MIRI.
The CEERS NIRCam coverage includes \jf{F115W}, \jf{F150W},
\jf{F200W}, \jf{F277W}, \jf{F356W}, \jf{F410M}, and \jf{F444W},
with $\texp\approx2.8$ks in each ($\texp\approx5.7$ks in \jf{F150W}).
The NIRCam WFSS uses \jf{F356W}. MIRI exposures include
\jf{F1000W}, \jf{F1280W}, \jf{F1500W}, \jf{F1800W}, and \jf{F2100W}.
NIRSpec coverage in CEERS features the medium resolution ($R\sim1000$)
grating with wavelength coverage $\lambda\approx0.97-5.14\um$ ($\texp\approx2.9$ks),
and $R\sim100$ prism observations with $\mab\approx26.5$ continuum sensitivity. 
An approximate exposure map of the CEERS \jf{F200W} imaging
is provided in Figure \ref{fig:jwst}.

Additional relevant DD-ERS programs include \jp{TEMPLATES: Targeting Extremely Magnified Panchromatic Lensed Arcs and Their Extended Star formation} \citep[DD-ERS 1355;][]{templates1355}.

\subsubsection{Guaranteed Time Observations}
\label{sec:gto}

Each of the \JWST{} instrument teams and interdisciplinary
scientists received Guaranteed Time Observations (GTO). The
GTO program covers an extremely wide range of science,
well beyond the \reion{} focus of this review, including
the interstellar medium, exoplanet research, and
stellar population investigations. The GTO programs
for extragalactic science feature incredibly powerful
combinations of filter arrays, spectroscopic modes,
depths, and parallel observations. Given the sensitivity
of \JWST{} to distant galaxies, nearly all of the
extragalactic GTO surveys are relevant for
studying galaxies during \reion{}.

The NIRCam-NIRSpec joint
Guaranteed Time Observations program called
\JWST{} Advanced Deep Extragalactic Survey (JADES) will 
survey GOODS-North and GOODS-South with a wide
range of NIRCam, MIRI, and NIRSpec observations.
The design of the GOODS-S survey involves a series of
NIRSpec prime - NIRCam parallel observations, followed
by NIRCam prime - MIRI parallel imaging, and concluding
with another set NIRSpec prime - NIRCam parallel exposures.
There are two characteristic NIRCam exposure depths for JADES
wide-band imaging,
including deep imaging at $\mab\approx29.8-30.35$ over $\sim46$~\sqarcmin{}
and medium imaging at $\mab\approx29.0-29.5$ over $\sim$290~\sqarcmin{}.
NIRCam images in JADES all include short-wavelength filters
\jf{F090W}, \jf{F115W}, \jf{F150W}, and \jf{F200W}, and 
long-wavelength filters \jf{F277W}, \jf{F356W}, \jf{F410M}, and \jf{F444W}.
Additional medium band images in \jf{F335M} are acquired
over subsets of the field to $\approx0.8$AB lower sensitivity, and we note that the
\jf{F410M} and \jf{F444W} images 
are shallower than the other broad-bands by $\sim0.2-0.5$AB
(but can be combined to match the deep depths).
NIRSpec observations are acquired with the prism, 
medium resolution spectroscopy with \jf{F070LP}, \jf{F170LP}, and \jf{F290LP},
and high-resolution spectroscopy with \jf{F290LP}.
Figure \ref{fig:jwst} shows approximate \jf{F200W} exposure maps for the JADES GOODS-N
and GOODS-S regions, where the relative subfield locations will depend on the
actual position angle of the observations.

The JADES GOODS-S observations are acquired through several
GTO Program IDs. Initial NIRSpec observations \citep[GTO 1210; 74.2h;][]{jades1210}
are acquired
in a single region targeting known objects near the HUDF, using prism,
medium, and high-resolution spectroscopy. In parallel,
deep NIRCam images are acquired at a single location. 
A compact NIRCam mosaic of deep and medium exposures is then
created \citep[GTO 1180; 355.7h;][]{jades1180}. 
A set of medium NIRSpec exposures near the HUDF are acquired,
along with medium NIRCam parallels arranged into a mosaic.
A set of deep and medium NIRCam prime images then completes the
contiguous mosaic, while exposing in MIRI \jf{F777W} and \jf{1280W}.
Two later sets of additional NIRSpec pointings then visit
the NIRCam mosaic, exposing the medium \citep[GTO 1286; 149.3h;][]{jades1286}
and deep \citep[GTO 1287; 74.2h;][]{jades1287} regions while imaging with
NIRCam in parallel. Note that Figure \ref{fig:jwst}
does not show the parallel NIRCam fields from GTO 1286, which
can be arranged into a rough mosaic in GOODS-S depending on when the
field is observed.

In JADES GOODS-N \citep[GTO 1181; 152.8h;][]{jades1181}, a region will be exposed
with NIRCam prime with MIRI parallel (\jf{F770W} and \jf{F1280W})
to medium depth. Initial NIRSpec pointings targeted from HST
sources in GOODS-N will be
acquired with NIRCam in parallel to continue the compact medium
depth mosaic with the same filters. The NIRSpec pointings
will use the prism and the medium resolution disperser
with \jf{F070LP}, \jf{F170LP}, and \jf{F290LP}.
Sources from the initial NIRCam GOODS-N area
will then be revisited with NIRSpec pointings that additionally
include high-resolution spectroscopy in \jf{F290LP}.
NIRCam parallels will be taken in a nearby region of 
CANDELS GOODS-N to medium depth. The NIRCam images in GOODS-N
cover the HDF-N region.

\begin{figure}[h]
\includegraphics[width=\linewidth]{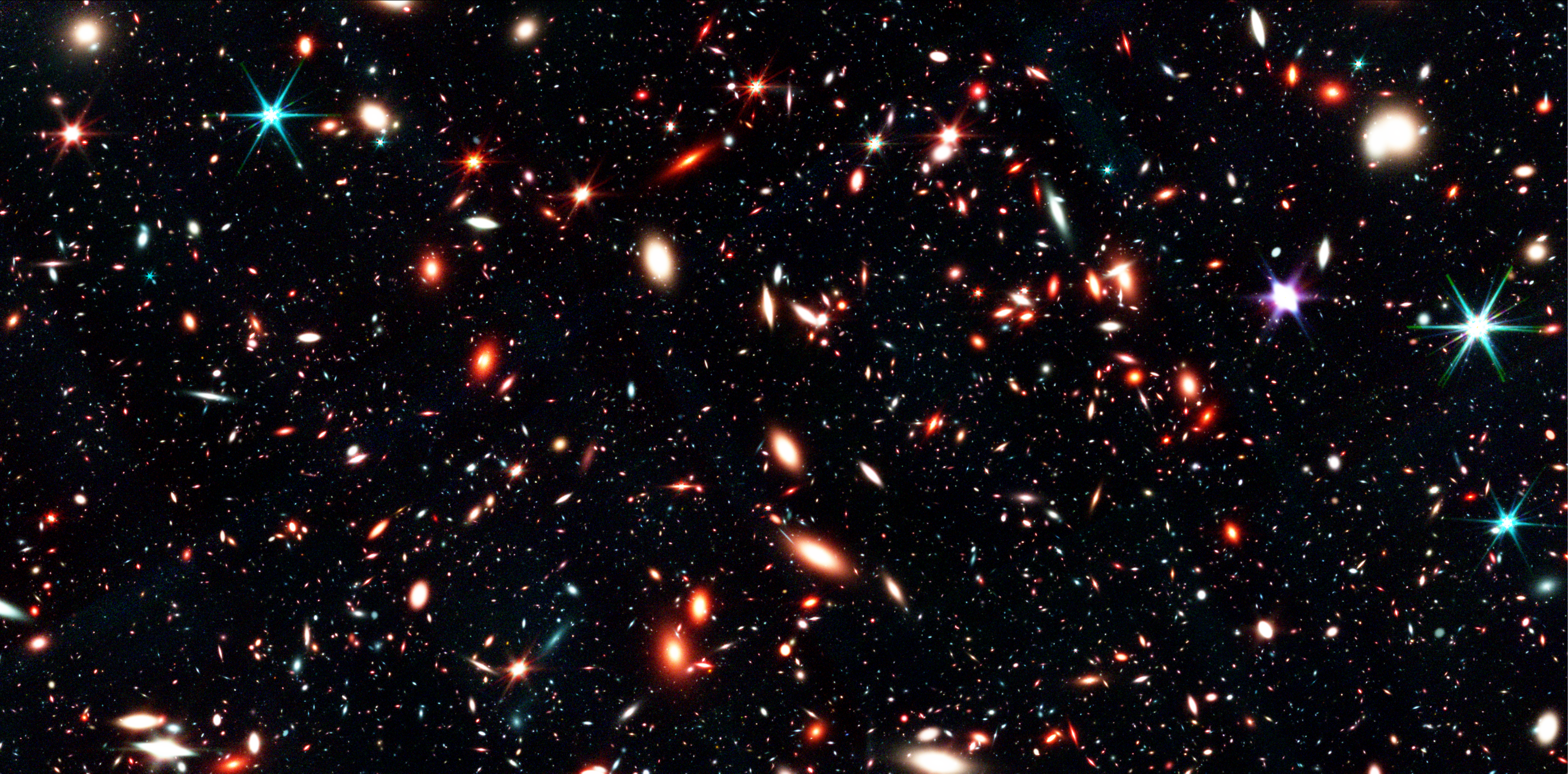}
\caption{Zoomed region of a simulation of the JADES Deep
imaging region in GOOD-S.
Shown is a \jf{F090W}-\jf{F115W}-\jf{F444W} false-color image created from synthetic data
produced
with the Guitarra \citep{willmer2020a} image simulator and
processed with the JADES mosaicing and analysis pipeline.
Adapted from \citet{williams2018a} with permission from the
author. JADES Deep NIRCam imaging will reach $f_\nu\approx29.4-30.35$ AB magnitude limits ($5\sigma$) for the $\lambda\approx0.90-4.4\um$
broadband filters.}
\label{fig:jades}
\end{figure}

As part of the GTO team efforts, the JADES collaboration
has been performing detailed simulations of their
survey data. Using the JAGUAR synthetic galaxy catalog
\citet{williams2018a} and the Guitarra image simulation
code \citep{willmer2020a}, the NIRCam GTO JADES team
has produced an end-to-end
model of the individual ramps, distortion, mosaicing,
pixel-level error propagation, detection, and photometry
for JADES. The NIRSpec GTO JADES team then performs
spectroscopic source selection and
simulated spectroscopy. Figure \ref{fig:jades} shows
a false color visualization of a simulated \jf{F090W}-\jf{F115W}-\jf{F444W}
multiband image (JADES Collaboration, priv. comm.),
adapted from \citet{williams2018a}.
The depth achievable by \JWST{} is illustrated by the
richness of the structures in this simulated
image, which corresponds to
a $5-\sigma$ sensitivity of $\mab=30.35$ in \jf{F115W}.

There are two MIRI programs targeting the Hubble Ultra
Deep Field. Program GTO 1207 \citep[89.7h;][]{miri1207,rieke2019a}
will observe the HUDF region with a $\approx$30~\sqarcmin{}
mosaic using all the MIRI imaging bands, and follow up
with $R=1000$ NIRSpec spectroscopy at $\lambda=0.97-5.14\um$.
Program GTO 1283 \citep[60.9h;][]{miri1283} will image the HUDF 
with a single pointing in \jf{F560W} for 60 hours, reaching $\mab\sim28.3$ ($SNR=4$).
Parallel observations will be acquired with NIRCam (\jf{F115W}, \jf{F277W}, \jf{F356W}) for
40 hours and NIRISS (\jf{F115W}, \jf{F150W}, and \jf{F200W} at $R\sim150$)
for 20 hours in nearby regions of GOODS-S.

Additional programs related to \reion{} include the Interdisciplinary
Scientist GTO Programs (\citealt[][GTO 1176]{windhorst1176}, and \citealt[][GTO 1243]{lilly1243}),
\jp{CANUCS: The CAnadian NIRISS Unbiased Cluster Survey} \citep[NIRISS; GTO 1208;][]{willott1208},
the \jp{Cosmic Re-ionization, Metal Enrichment, and Host Galaxies from Quasar Spectroscopy}
\citep[NIRSpec; GTO 1222;][]{nirspec1222}, and the NIRSpec WIDE survey \citep[GTO 1211-1215;][]{nirspec1211,nirspec1213,nirspec1214,nirspec1215}.

\subsubsection{GO Programs}
\label{sec:go}

A series of \JWST{} Guest Observer programs
will pursue extremely ambitious \reion{}
science. There are roughly three
dozen GO extragalactic programs that
explore distant galaxies near reionization.
Below we discuss five and note several
others.

COSMOS-Webb \citep[GO 1727; 207.8h;][]{cosmoswebb1727}
is the largest targeted extragalactic GO program,
both in time and area, covering
approximately $\sim$0.6deg$^{2}$ with
NIRCam and $\sim$0.2deg$^{2}$ with MIRI.
The NIRCam images expose
\jf{F115W}, \jf{F150W}, \jf{F227W},
and \jf{F444W} to $\mab\approx27.4-28.0$, while
MIRI exposes in parallel \jf{F770W}
to $\mab\approx24.3-24.5$. Figure \ref{fig:jwst}
shows the \jf{F115W} exposure map for COSMOS-Webb.
The extreme area will enable COSMOS-Webb to
discover rare, bright objects during \reion{},
while the contiguous mosaic will allow for
detailed clustering of high-redshift galaxies.

The Public Release IMaging for Extragalactic Research
\citep[GO 1837; 188.1h;][]{primer1837} survey will observe
the HST CANDELS regions in COSMOS and UDS. The
designs of the two mosaics are carefully orchestrated
to provide substantial overlap between the MIRI
prime imaging and the NIRCam parallel imaging.
The NIRCam (MIRI) mosaics are 144~\sqarcmin{} (112~\sqarcmin{})
in COSMOS and 234~\sqarcmin{} (125~\sqarcmin{}) in UDS, and
each field has a NIRCam-MIRI overlap area of more than
100~\sqarcmin{}. The PRIMER NIRCam images use
\jf{F090W}, \jf{F115W}, \jf{F150W}, and \jf{F200W}),
\jf{F277W}, \jf{F356W}, \jf{F410M}, and \jf{F444W},
and will reach $\mab\approx28.5-29.5$ (\jf{F200W}).
The MIRI images
use \jf{F770W} and \jf{F1800W} to reach 
$\mab\approx25.6$ in \jf{F770W}.
Figure \ref{fig:jwst} shows approximate
exposure maps for the PRIMER \jf{F200W} imaging in 
COSMOS and the UDS, constructed from the public APT files.

The Webb Deep Extragalactic Exploratory Public (WDEEP) Survey
\citep[GO 2079; 121.7h;][]{wdeep2079} uses extremely sensitive
observations with NIRISS in the HUDF and NIRCam on HUDF
Parallel 2. 
The WDEEP NIRISS observations will reach line sensitivities of
$f_\nu\approx10^{-18}$ ergs s$^{-}$ cm$^{-2}$ in \jf{F115W}
, \jf{F150W}, and \jf{F200W} grism observations,
covering the extremely deep HST ACS imaging in the HUDF. 
WDEEP can
leverage the ultradeep \jf{F814W} HST imaging in HUDF Parallel 2
from the UDF12 project \citep{ellis2013a} as a
veto for its deep NIRCam \jf{F115W}, \jf{F150W}, \jf{F200W},
\jf{F356W}, and \jf{F444W} images in
conducting high redshift galaxy searches to $\mab\sim30.6-30.9$.
The NIRCam imaging area is slightly larger than a single pointing,
and an approximate exposure map for WDEEP is shown in Figure \ref{fig:jwst}.

The First Reionization Epoch Spectroscopic
COmplete Survey \citep[FRESCO; GO 1895; 53.1h;][]{fresco1895} will
conduct \jf{F444W} NIRCam grism observations in two mosaics,
each covering a 60~\sqarcmin{} region of GOODS-N or GOODS-S.
FRESCO will reach a line sensitivity of roughly $f_{\nu}\approx3\times10^{-18}$
ergs s$^{-1}$ cm$^{-2}$, allowing for the identification of
more than 300 galaxies at $z>7$. By using the grism,
FRESCO can attempt to capture the rest-frame optical emission
of high-redshift galaxies with an effectively flux-limited
selection, which will provide an interesting view into the
star formation and internal ionizing radiation fields of 
distant sources.

The Ultra-deep NIRCam and NIRSpec Observations Before the
Epoch of Reionization \citep[UNCOVER; GO 2561; 68.2h;][]{uncover2561} will image the
gravitational lensing cluster Abell 2744
in NIRCam \jf{F115W}, \jf{F150W}, \jf{F200W}, 
\jf{F277W}, \jf{F356W}, \jf{F410M}, and \jf{F444W}
to $\mab\approx29.8$), followed by extremely
deep $\texp\approx19$h NIRSpec prism spectroscopy.
The program will obtain deep parallel NIRISS and
NIRCam imaging in a nearby field. UNCOVER will establish
one of the intrinsically deepest observations with \JWST{},
with the ability to probe further down the high-redshift
luminosity function than larger area surveys.

\JWST{} will conduct many other notable GO programs
relevant for \reion{}. We briefly mention an incomplete
list here. Two pure parallel programs,
PASSAGE \citep[GO 1571; 591h;][]{malkan1571} with NIRISS
and PANORAMIC \citep[GO 2514; 150h;][]{williams2514} with NIRCam
will observe a large area and potentially find many
bright, high-redshift galaxies.
The spectroscopic programs AURORA \citep[GO 1914; 63.8h;][]{shapley1914}
and CECILIA \citep[GO 2593; 39.2h;][]{strom2593} will provide
direct-method calibrations of high-redshift galaxy metallicities,
which will inform models for the ionizing photon production efficiency.
The UDF Medium Band survey \citep[GO 1963; 20.4h;][]{williams1963} will
provide NIRCam imaging in the HUDF with \jf{F182M},
\jf{F210M}, \jf{F430M} \jf{F460M}, and \jf{F480M}, along
with NIRISS \jf{F430M} and \jf{F480M} imaging in parallel.
At least two programs will perform spectroscopy on 
\Lyc{} emitting galaxies, including NIRSpec IFU observations
 \citep[GO 1827; 24.1h;][]{laces1827}
and NIRSpec medium-resolution spectroscopy \citep[GO 1869; 73.5h;][]{schaerer1869},
both observing $z\sim3$ strong-line emitters.

\section{SUMMARY}
\label{sec:summary}

Over the last 5-10 years, our knowledge of galaxy formation
during \reion{} has advanced dramatically. The discovery
of photometric candidates at $z\sim6-10$ with \HST{}
and other facilities has revealed a substantial population
of galaxies forming just a few hundred million years
after the Big Bang. The 
spectroscopic confirmation of very distant galaxies out to
$z\sim8-10.5$ has now solidified the reality of these
early objects. With the coming of \JWST{}, which will find
more and fainter galaxies during reionization and 
measure their rest-frame optical spectra, we will begin
to understand the physics behind the role galaxies
play in \reion{}. Instead of wondering whether galaxies
dominated \reion{}, research questions will focus on 
the physical mechanisms by which galaxies produce and release
Lyman continuum radiation and the properties of the
galaxy population. \JWST{} will help us answer how
galaxies were able to rewrite cosmic
history by reionizing the universe.

\begin{summary}[SUMMARY POINTS]
\begin{enumerate}
\item Deep, near-infrared surveys have now established that galaxies primarily reionized the universe,
and radiation from accretion onto supermassive black holes in AGN plays a less dominant role.
\item The ultraviolet luminosity density declines from $z\sim6-10$, but the continued decline beyond $z\sim10$ is still unconfirmed.
\item Our understanding of the ability for stellar populations to produce efficiently ionizing photons has substantially improved, guided by both population synthesis modeling and observational evidence for strong nebular emission in distant galaxies.
\item The escape of Lyman continuum photons from galaxies during \reion{} is complex and remains a challenge for understanding the detailed physics of the reionization process.
\item The discovery of very bright, distant galaxies during the epoch of reionization may rebalance conclusions about the relative importance of bright versus faint galaxies in \reion{}.
\item Redshift confirmation of distant galaxies at redshifts $z\sim8-10$ has established the reality of star-forming systems less than 500Myr after the Big Bang.
\item Stellar mass function constraints at high-redshift are improving, and the presence of evolved stars in early galaxies provide circumstantial evidence for extended star formation to $z\sim15$ that \JWST{} could potentially verify.
\end{enumerate}
\end{summary}

\begin{issues}[FUTURE ISSUES]
\begin{enumerate}
\item \JWST{} will provide a revolutionary facility for studying \reion{} and will execute many exciting programs in its first year of operations.
\item The complexity of the \JWST{} data, including in the wide range of sensitive photometric filters, the power of wide-field slitless spectroscopy, and the new advance of highly multiplexed infrared spectroscopy in space, will present a rewarding challenge for the astronomical community.
\item \JWST{} can discover and potentially confirm galaxies out to extraordinarily high redshifts, but it remains unclear, given the decline of the high-redshift luminosity density, how many galaxies beyond $z\sim10$ will be newly revealed.
\item The ability to measure rest-frame optical spectra, estimate ionizing photon production through nebular emission, and constrain evolved stellar populations in galaxies at redshifts $z>8$ will place an increased emphasis on the detailed modeling of early galaxy evolution.\
\item Questions about the evolution of the Lyman continuum escape fraction, how it depends on galaxy properties including the efficiency of ionizing photon production, will remain an area of active research.
\item Understanding and interpreting the detailed spectra of distant galaxies will provide a rich area for exploring \reion{} with \JWST{}, including studies of the excitation of near UV metal lines, the propagation of \Lya, the connection between interstellar absorption and Lyman continuum escape, direct calibration of metallicity indicators, and the connection between metallicity and ionizing photon production.
\end{enumerate}
\end{issues}

\section*{DISCLOSURE STATEMENT}
The author is not aware of any affiliations, memberships, funding, or financial holdings that
might be perceived as affecting the objectivity of this review. 

\section*{ACKNOWLEDGMENTS}
BER thanks Sandy Faber for carefully reading the manuscript
and providing detailed comments that improved this review.
Thanks to Pascal Oesch and Christina Williams for permission
to use or adapt their figures, and Rychard Bouwens for providing
the data in Figure \ref{fig:lf}. Thanks also to the JADES 
collaboration for providing Figure \ref{fig:jades}, with
special thanks to Sandro Tacchella, Christopher Willmer, and
Marcia Rieke. Also, thanks to Anton Koekemoer for providing
the updated ATP file for COSMOS-Webb.
BER acknowledges support from NASA contract NNG16PJ25C
and grant 80NSSC18K0563.

\begin{table}[h]
\tabcolsep7.5pt
\caption{\citet{schechter1976a} Luminosity Function $\phi(\MUV)$ Parameters}
\label{tab:schechter}
\begin{center}
\begin{tabular}{@{}l|c|c|c|c@{}}
\hline
 & $\phi^{\star}$& $M^{\star}$&&\\
Redshift&{(}Mpc$^{-3}$ mag$^{-1}$) &{(}mag) &$\alpha$ &Reference\\
\hline
$z\sim6$  &$5.1^{+1.2}_{-1.0}\times10^{-4}$ &$-20.93\pm0.09$ &$-1.93\pm0.08$ & \citet{bouwens2021a}\\
$z\sim7$  &$1.9^{+0.8}_{-0.6}\times10^{-4}$ &$-21.15\pm0.13$ &$-2.06\pm0.11$ & \citet{bouwens2021a}\\
$z\sim8$  &$9^{+9}_{-5}\times10^{-5}$       &$-20.93\pm 0.28$ &$-2.23\pm0.20$ & \citet{bouwens2021a}\\
$z\sim9$  &$2.1^{+1.4}_{-0.9}\times10^{-5}$ &$-21.15 (fixed)$ &$-2.33\pm0.19$ & \citet{bouwens2021a}\\
$z\sim10$ &$4.2^{+4.5}_{-2.2}\times10^{-6}$ &$-21.19 (fixed)$ &$-2.38\pm0.28$ & \citet{oesch2018a}\\
\hline
\end{tabular}
\end{center}
\end{table}

\begin{table}[h]
\tabcolsep7.5pt
\caption{Double Power-Law Luminosity Function $\phi_{\mathrm{DPL}}(\MUV)$ Parameters}
\label{tab:dpl}
\begin{center}
\begin{tabular}{@{}l|c|c|c|c|c@{}}
\hline
 & $\phi^{\star}_{\mathrm{DPL}}$& $M^{\star}_{\mathrm{DPL}}$&&&\\
Redshift&{(}Mpc$^{-3}$ mag$^{-1}$) &{(}mag) &$\alpha$ &$\beta$ &Reference\\
\hline
$z\sim6$ & $3.02^{+0.70}_{-0.45}\times10^{-4}$ &$-21.03^{+0.09}_{-0.03}$ & $-2.08^{+0.07}_{-0.06}$ &$-4.57^{+0.09}_{-0.10}$ & \citet{harikane2021a}\\
$z\sim7$ & $8.91^{+2.31}_{-1.83}\times10^{-4}$ &$-20.12^{+0.12}_{-0.10}$ & $-1.89^{+0.10}_{-0.09}$ &$-3.81^{+0.10}_{-0.13}$ & \citet{harikane2021a}\\
$z\sim8$ & $4.83^{+2.25}_{-2.25}\times10^{-4}$ &$-19.80\pm0.26$          &$-1.96\pm0.15$           &$-3.98\pm0.14$          & \citet{bowler2020a}\\
$z\sim9$ & $2.85^{+1.39}_{-1.39}\times10^{-4}$ &$-19.67\pm0.33$          &$-2.1$ (fixed)           &$-3.75\pm0.22$          & \citet{bowler2020a}\\
\hline
\end{tabular}
\end{center}
\end{table}

\begin{table}[h]
\tabcolsep5.5pt
\caption{Luminosity Density $\rho_{\mathrm{UV}}$}
\label{tab:rhoUV}
\begin{center}
\begin{tabular}{@{}l|c|c|c|c|c|c@{}}
\hline
        & $\log_{10}\rho_{\mathrm{UV}}\,^{\rm a}$ &$\rho_{\mathrm{UV}}(>L^{\star})/$&$\rho_{\mathrm{UV}}(>0.1L^{\star})/$& $\log_{10} \rho^{\mathrm{DPL}}_{\mathrm{UV}}\,^{\rm a}$&$\rho^{\mathrm{DPL}}_{\mathrm{UV}}(>L^{\star})/$&$\rho^{\mathrm{DPL}}_{\mathrm{UV}}(>0.1L^{\star})/$\\
Redshift&$(>0.01L^{\star})$&
$\rho_{\mathrm{UV}}(>0.01L^{\star})$&
$\rho_{\mathrm{UV}}(>0.01L^{\star})$&
$(>0.01L^{\star})$&
$\rho^{\mathrm{DPL}}_{\mathrm{UV}}(>0.01L^{\star})$&
$\rho^{\mathrm{DPL}}_{\mathrm{UV}}(>0.01L^{\star})$\\
\hline
$z\sim6$ & $26.25^{\rm b}$ &$0.066$ & $0.49$&$26.27$ & $0.048$ & $0.45$ \\
$z\sim7$ & $26.05$ &$0.046$ & $0.42$&$26.20$ & $0.106$ & $0.57$ \\
$z\sim8$ & $25.83$ &$0.027$ & $0.32$&$25.87$ & $0.084$ & $0.52$ \\
$z\sim9$ & $25.41$ &$0.019$ & $0.28$&$25.72$ & $0.068$ & $0.44$ \\
$z\sim10$& $24.80$ &$0.016$ & $0.25$&\ldots  & \ldots & \ldots \\
\hline
\end{tabular}
\end{center}
\begin{tabnote}
$^{\rm a}${(}erg s$^{-1}$ Hz$^{-1}$ Mpc$^{-3})$;
$^{\rm b}$ To convert $\rhoUV$ to a star formation
rate density $\dot{\rho}_{\star}$ [$\Msun$ yr$^{-1}$ Mpc$^{-3}$],
a useful approximation is $\log_{10} \dot{\rho}_{\star}\approx\log_{10} \rhoUV - 28.1$.
This conversion assumes a 100 Myr-old, 
constant star formation rate stellar population without dust \citep{eldridge2017a}.
\end{tabnote}
\end{table}

\end{document}